%% file: aa6820.tex
\documentclass{aa}  
\usepackage{amsmath}
\usepackage{graphicx}
\usepackage{natbib}
\usepackage{txfonts}

\usepackage{ulem}
\def\new#1 {#1 }
\def\old#1 {}

\include{defs}

\def\radex{\texttt{RADEX}}

\begin{document}
\title{A computer program for fast non-LTE analysis of interstellar line spectra}

\subtitle{With diagnostic plots to interpret observed line intensity ratios}

\author{F. F. S. van der Tak\inst{1,2} \and J. H. Black\inst{3}
        \and F. L. Sch{\"o}ier\inst{4} \and D. J. Jansen\inst{5}
        \and E. F. van Dishoeck\inst{5}}

\titlerunning{Fast non-LTE analysis of interstellar line spectra}
\authorrunning{Van der Tak et al.}

\institute{Netherlands Institute for Space Research (SRON), Landleven 12, 9747 AD Groningen, The Netherlands
              \\ \email{vdtak@sron.rug.nl} 
      \and Max-Planck-Institut f\"ur Radioastronomie, Auf dem H\"ugel 69, 53121 Bonn, Germany
      \and Onsala Space Observatory, Chalmers University of Technology, 43992 Onsala, Sweden
      \and Stockholm Observatory, AlbaNova University Center, 10691 Stockholm, Sweden
      \and Leiden University Observatory, P.O. Box 9513, 2300 RA Leiden, The Netherlands}

\date{Received 27 November 2006; accepted 27 March 2007}

\abstract
{}
{The large quantity and high quality of modern radio and infrared line
  observations require efficient modeling techniques to infer physical and
  chemical parameters such as temperature, density, and molecular abundances.}
{We present a computer program to calculate the intensities of atomic and
  molecular lines produced in a uniform medium, based on statistical equilibrium
  calculations involving collisional and radiative processes and including
  radiation from background sources. Optical depth effects are treated with an
  escape probability method. The program is available on the World Wide
  Web at \texttt{ http://www.sron.rug.nl/$\sim$vdtak/radex/index.shtml }. The
  program makes use of molecular data files maintained in the Leiden Atomic and
  Molecular Database (LAMDA), which will continue to be improved and expanded. }
{The performance of the program is compared with more approximate and with more
  sophisticated methods. An Appendix provides diagnostic plots to estimate
  physical parameters from line intensity ratios of commonly observed
  molecules.}
{This program should form an important tool in analyzing observations from
  current and future radio and infrared telescopes.}


\maketitle

\section{Introduction}

Observations of spectral lines at radio, (sub)millimeter and infrared
wavelengths are a powerful tool to investigate the physical and chemical
conditions in the dilute gas of astronomical sources where thermodynamic
equilibrium is a poor approximation \citep[e.g.,][]{genzel:crete1,black:korea}.
To extract astrophysical parameters from the data, the excitation and optical
depth of the lines need to be estimated, for which various methods may be
used, depending on the available observations \citep{vdishoeck:creteII,vdtak:catania}.

If only one or two lines of a molecule\footnote{In this paper, the term
  `molecule' includes mono-atomic and ionic species.} have been observed, the
excitation must be deduced from observations of other species or from
theoretical considerations.  An example is the assumption that the excitation
temperature equals the kinetic temperature, a case known as Local Thermodynamic
Equilibrium (LTE) which holds at high densities.

If many lines have been observed, \old{such as in a spectral line survey,} a popular
method is the `rotation diagram', also called `Boltzmann plot' or `population
diagram' \citep[e.g.,][]{blake:orion,helmich:w3,goldsmith:population}. 
This method \old{assumes that the lines are optically thin and} describes the
excitation by a single temperature, \new{obtained by a fit to the line
  intensities as a function of upper level energy}. Provided that beam sizes are
similar and optical depths are low, or that appropriate corrections are made,
this method yields estimates of the excitation temperature and column density of
the molecule.
The excitation temperature approaches the kinetic temperature in the
high-density limit, but generally depends on both kinetic temperature and volume density.
\new{Spectral line surveys are often analyzed with rotation diagrams, although
  more advanced methods are also used \citep{helmich:survey,comito:survey}.}

More sophisticated methods retain the assumption of a local excitation, but
solve for the balance of excitation and de-excitation rates from and to a given
state, the so-called statistical equilibrium (SE). The best-known methods of
this type are the escape probability method and the Large Velocity Gradient
(LVG) method (\citealt{sobolev:book}; \citealt{dejong:collapse}; \citealt{goldreich:lvg}).
These `intermediate-level' methods require knowledge of
molecular collisional data, whereas the previous `basic-level' methods only
required spectroscopic \new{and dipole moment} information. This extra requirement limits the use of
these methods to some extent, because collisional data do not exist for all
astrophysically relevant species.
The advantage is that column density, kinetic temperature and volume density can
be constrained, if accurate collision rates are known.
\new{As with rotation diagrams, this method can be used to compute synthetic
  spectra to be compared with data with a $\chi^2$ statistic \citep{jansen:thesis,leurini:ch3oh}.}

The most advanced methods drop the local approximation and solve for
the intensities (or the radiative rates) as functions of depth into
the cloud, as well as of velocity. Such methods are usually of the
Accelerated Lambda Iteration (ALI) or Monte Carlo (MC) type, although
hybrids also exist. The performance and convergence of such programs
have recently been tested by \citet{zadelhoff:benchmark}.  Using such
programs one can constrain temperature, density, and velocity
gradients within sources
\citep[e.g.,][]{vdtak:gl2591,tafalla:starless,jakob:dr21}, 
and, if enough observations are available, even molecular abundance profiles
\citep[e.g.,][]{vdtak:ch3oh,schoeier:16293,maret:meth}, especially when coupled to
chemical networks \citep[e.g.,][]{doty:16293,evans:b335,goicoechea:horsehead}. 

This paper presents the public version of a radiative transfer code at the
`intermediate' level.  The assumption of a homogeneous medium limits the number
of free parameters and makes the program a useful tool in rapidly analyzing a
large set of observational data, in order to provide constraints on physical
conditions, such as density and kinetic temperature \citep{jansen:thesis}.
The program can be used for any molecule for which collisional
rate coefficients are available. The input format for spectroscopic and
collisional data is that of the LAMDA database \citep{schoeier:lamda}\footnote{\tt
  http://www.strw.leidenuniv.nl/$\sim$moldata} where an on-line calculator for
molecular line intensities\footnote{In this paper, the `strength' of a
line is an intrinsic quantity related to its transition dipole monent,
while its `intensity' is an observable related to the emission from a
celestial object}, based on our program, can also be found\footnote{\tt
  http://www.sron.rug.nl/$\sim$vdtak/radex.php}.

The paper is set up as follows. Section~\ref{s:formal} describes the radiative
transfer formalism and introduces our notation of the key quantities.
Section~\ref{s:radex} describes the formalism which the program actually uses, and
discusses its implementation.  Section~\ref{s:compare}
compares the results of the program to those of other programs.  The paper
concludes in \S~\ref{s:concl} with suggested future directions of astrophysical
radiative transfer modeling.


%
\section{Radiative transfer and molecular excitation}
\label{s:formal}
This section summarizes the formalism to analyze molecular line observations
which our program adopts.  For more detailed discussions of radiative transfer
see, e.g., \citet{cannon:book} or \citet{rybicki:lightman}.

\subsection{Basic formalism}
Describing the transfer of radiation requires a quantity which is conserved
along its path as long as no local absorption or emission processes take place,
and which includes the direction of travel. The quantity that satisfies this
requirement is the specific intensity $I_\nu$, defined as the amount of energy
passing through a surface normal to the path, per unit time, surface, bandwidth
(measured here in frequency units), and solid angle. 
The transfer equation for radiation propagating a distance $ds$ can then
be written as
\begin{equation}
\frac{dI_\nu}{ds} = j_\nu - \alpha_\nu\,I_\nu,
\end{equation}
where $j_\nu$ and $\alpha_\nu$ are the local emission and extinction 
coefficients, respectively. 
\old{A useful quantity for subsequent use is the mean intensity}
%
%
\old{i.e., the specific intensity $I_{\nu}$ integrated over solid angle $d\Omega$ and
averaged over all directions.}
\old{For radiation theory, an important and useful quantity known as the source
function is defined by }
\new{The two terms on the right-hand side may be combined into the source
  function, defined by }
\begin{equation}
\label{sourcefuncdef}
S_\nu \equiv \frac{j_\nu}{\alpha_\nu}.
\end{equation}
Writing the transport equation in its integral form and defining the optical
depth, $d\tau_{\nu} \equiv \alpha_{\nu}\,ds$, measured 
along the ray\footnote{In this paper, `ray' denotes a light path of
  infinitesimally small width, and `beam' is a collection of rays within a
  finite-sized solid angle.} one arrives at
\begin{equation}
\label{transfeq}
I_\nu = I_\nu\,(0) e^{-\tau_{\nu}} + 
\int^{\tau_{\nu}}_0\,S_\nu(\tau_{\nu}^{\prime})\,e^{-(\tau_{\nu} - \tau^{\prime}_{\nu})}
\,d\tau^{\prime}_{\nu}, 
\end{equation}
where $I_\nu$ is the radiation emerging from the medium and $I_\nu(0)$ is the
`background' radiation entering the medium.  

The above equations hold both for continuum radiation, which is emitted over a
large bandwidth, and for spectral lines, which arise when the local emission and
absorption properties change drastically over a very small frequency interval,
due to the presence of molecules. From this point the discussion will focus on
bound-bound transitions within a multi-level molecule consisting of $N$ levels
with spontaneous downward rates $A_{ul}$, Einstein coefficients for stimulated
transitions $B_{ul}$ and $B_{lu}$, and collisional rates $C_{ul}$ and $C_{lu}$,
between upper levels $u$ and lower levels $\ell$.

The rate of collision is equal to
\begin{equation}
C_{ul} = n_{\mathrm{col}} \gamma_{ul},
\end{equation}
where $n_{\mathrm{col}}$ is the number density of the collision
partner (in \ccm) and $\gamma_{ul}$ is the downward collisional rate coefficient (in
cm$^3$\,s$^{-1}$).  The rate coefficient is the Maxwellian average of
the collision cross section, $\sigma$,
\begin{equation}
\label{ratecoeff}
\gamma_{ul} = \left(\frac{8kT_{\rm kin}}{\pi\mu}\right)^{-1/2}\left(\frac{1}{kT_{\rm kin}}\right)^2\int \sigma E e^{-E/kT_{\rm kin}} dE,
\end{equation}
where $E$ is the collision energy, $k$ is the Boltzmann constant, \tkin\ is the
kinetic temperature, and $\mu$ is the reduced mass of the system.  The upward
rates are obtained through detailed balance
\begin{equation}
\label{eq:dbalance}
\gamma_{lu} = \gamma_{ul} \frac{g_u}{g_l} e^{-h\nu/kT_{\mathrm{kin}}},
\end{equation}
where $g_i$ is the statistical weight of level $i$.

The local emission in transition $u$$\rightarrow$$l$ with laboratory 
frequency $\nu_{ul}$, can be expressed as 
\begin{equation}
\label{localem}
j_\nu = \frac{h\nu_{ul}}{4\pi}\,n_u\,A_{ul}\,\phi_\nu,
\end{equation}
where $n_u$ is the number density of molecules in level $u$ and 
$\phi_\nu$ is the frequency-dependent line emission profile.
The absorption coefficient reads
\begin{equation}
\label{tau}
\alpha_\nu = \frac{h\nu_{ul}}{4\pi} \left( n_l\,B_{lu}\,\varphi_\nu - 
n_u\,B_{ul}\,\chi_\nu \right),
\end{equation}
where $\phi_\nu$ and $\chi_\nu$ are the line profiles for absorption and 
stimulated emission (counted as negative extinction), respectively. 

From here on we assume complete angular and frequency redistribution of the
emitted photons, so that $\phi_\nu$$=$$\varphi_\nu$$=$$\chi_\nu$, which is
strictly only valid when collisional excitation dominates.  This assumption
allows the source function to be written as
\begin{equation}
\label{sourcefunc}
S_{\nu_{ul}} = \frac{n_u A_{ul}}{n_l B_{lu} - n_u B_{ul}} = \frac{2h\nu_{ul}^3}{c^2}
\left( \frac{g_u n_l}{g_l n_u} - 1 \right)^{-1}, 
\end{equation}
where we have used the Einstein relations.  It is common to introduce
an excitation temperature $T_{\rm ex}$ defined through the Boltzmann
equation
\begin{equation}
\label{Boltzmann}
\frac{n_u}{n_l}=\frac{g_u}{g_l}\exp\left[-(E_u-E_l)/kT_{\mathrm{ex}}\right],
\end{equation}
where $E_i$ is the energy of level $i$, such that 
%
$S_{\nu_{ul}} = B_{\nu}(T_{\mathrm{ex}})$, 
%
the specific intensity of a blackbody
radiating at $T_{\mathrm{ex}}$.

In the interstellar medium, the dominant line broadening mechanism is Doppler
broadening. Except in very cold and dark cloud cores, observed line widths are
much larger than expected from the kinetic temperature: this effect is commonly
ascribed to random macroscopic gas motions or `turbulence'. The result is a
Gaussian line profile
\begin{equation}
\label{lineprofile}
\phi_\nu = \frac{1}{\nu_{\mathrm D} \sqrt{\pi}} \exp \left[ - \left( \nu - \nu_{ul} -
\vec{v}\cdot\vec{n}\,\frac{\nu_{ul}}{c} \right)^2 / \nu_{\mathrm D}^2 \right],
\end{equation}
where $\nu_{\mathrm D}$ is the Doppler width, $\vec{v}$ is the velocity vector
of the moving gas at the position of the scattering, $\vec{n}$ is a unit vector
in the direction of the propagating beam of radiation, and $c$ is the speed of
light. \new{The Doppler width is the $1/e$ half-width of the profile, equal to $\Delta V /
  2\sqrt{\ln 2}$ where $\Delta V$ is its full width at half-maximum}.

If the level populations $n_i$ are known, the radiative transfer equation can be
solved exactly. In particular, under LTE conditions, knowledge of the kinetic
gas temperature $T_{\mathrm{kin}}$ allows the determination of $n_i$ by virtue
of the Boltzmann equation (Eq.~\ref{Boltzmann}).  For many interstellar and
circumstellar media, the density is too low to attain LTE, but statistical
equilibrium (SE) can often be assumed:
\begin{equation}
\label{SE}
\frac{dn_i}{dt} = 0 = \sum_{j\,\neq\,i}^{N} n_j P_{ji} - n_i\sum_{j\,\neq\,i}^{N}
P_{ij} = {\cal F}_i - n_i {\cal D}_i \;\;\; ,
\end{equation}
\new{where $P_{ij}$, the destruction rate coefficient of level $i$,
and its formation rate coefficient $P_{ji}$ are given by}
%
%
\begin{equation}
  \label{eq:rate}
  P_{ij} = 
\begin{cases}
A_{ij} + B_{ij} \bar{J_\nu} + C_{ij} & ( i > j ) \\
        B_{ij} \bar{J_\nu}  + C_{ij} & ( i < j ).
\end{cases}
\end{equation}

In Eq.~\ref{eq:rate},
\begin{equation}
\label{Sul}
B_{ij}\bar{J_\nu} = B_{ij}\int^{\infty}_0 J_\nu\,\phi(\nu)\,d\nu
\end{equation}
is the number of induced radiative (de-)excitations from state $i$ to state $j$
per second per particle in state $i$, and
\begin{equation}
\label{jnu}
J_\nu = \frac{1}{4\pi}\int I_\nu\,\mathrm{d}\Omega
\end{equation}
is the specific intensity $I_{\nu}$ integrated over solid angle $d\Omega$ and
averaged over all directions.
The SE equations thus include the effects of non-local radiation.


\new{This discussion assumes that} the state-specific rates of formation ${\cal
  F}_i$ [cm$^3$\,s$^{-1}$] and destruction ${\cal D}_i$ [s$^{-1}$] \old{can
  often be set to} \new{are} zero to ensure that the \old{excitation and}
radiative transfer \old{are} \new{is} solved independently of assumptions about
chemical processes.  \old{However,} \new{In general,} formation and destruction
processes should be included explicitly to be able to deal with situations in
which the chemical time scales are very short or the radiative lifetimes very
long. For example, the formation temperature (in ${\cal F}_i$) affects the
rotational excitation of \old{molecules like} C$_3$ \citep{roueff:c3} and
\new{the vibrational excitation of} \hh\ \new{\citep{black:h2,burton:h2pumping,takahashi:h2}}, \new{systems} for
which \old{rotational lines} \new{line radiation} only occurs as
slow electric quadrupole transitions. The rotational excitation of reactive ions
like CO$^+$ (\citealt{fuente:co+}; \citealt{black:faraday}) is also
sensitive to ${\cal F}_i$ and ${\cal D}_i$ because the rates of reactions with H
and \hh\ rival the inelastic collision excitation rates. Similar
considerations apply to the excitation of \hhhp\ in the Sgr~A region close to
the Galactic Center \citep{vdtak:london}, where electron impact excitation
competes with dissociative recombination.

\subsection{Molecular line cooling}
Once the radiative transfer problem has been solved and the level
populations are known, the cooling (or heating) from molecular line
emission can be estimated. Since the level populations contain all the
information of the radiative transfer, a general expression for the
cooling is obtained from considering all possible collisional
transitions
\begin{equation}
\Lambda = \sum_{i}n(i)\sum_{l}\sum_{u>l} ( n_l \gamma_{lu} - n_u \gamma_{ul} ) h\nu_{ul},
\end{equation}
where $i$ denotes a collision partner. The emissivity $\Lambda$ in
erg\,s$^{-1}$\,\ccm\ is defined to be positive for net cooling. From
Eq.~(\ref{eq:dbalance}), the collision rate coefficients $\gamma_{lu}$ and
$\gamma_{ul}$ are in detailed balance at the kinetic temperature; therefore it
is possible for net heating to occur ($\Lambda < 0$) in cases where the crucial
level populations have \txc$>$\tkin, owing to strong radiative excitation in a
hot external radiation field.

\subsection{Escape probability} 
\label{escape}

The difficulty in solving radiative transfer problems is the interdependence of
the molecular level populations and the local radiation field, requiring
iterative solution methods.
In particular, for inhomogeneous or geometrically complex objects, extensive
calculations with many grid points are required.
However, if only the global properties of an interstellar cloud are of interest,
the calculation can be greatly simplified through the introduction of a
geometrically averaged escape probability $\beta$, the probability that a photon
will escape the medium from where it was created. This probability depends only
on the optical depth $\tau$ and is related to the intensity within the medium,
\new{ignoring background radiation and any local continuum,}
through
\begin{equation}
\overline{J}_{\nu_{ul}} = S_{\nu_{ul}}(1-\beta).
\end{equation}

Several authors have developed detailed relations between $\beta$ and $\tau$ for
specific geometrical assumptions. Our program offers the user a choice of three
such expressions. 
The first is the expression derived for an expanding spherical shell, the
so-called Sobolev or large velocity gradient (LVG) approximation
(\citealt{sobolev:book}; \citealt{castor:escprob}; \citealt{elitzur:masers},
p.~42-44). \new{This method is also widely applied for moderate velocity
  gradients, to mimic turbulent motions.}
Our program uses the formula by \new{\citet{mihalas:book} and} \citet{dejong:lvg} for this geometry:

\begin{equation}
\beta_{\mathrm{LVG}} = \frac{1}{\tau}\int^{\tau}_0 e^{-\tau^{\prime}}d\tau^{\prime} = \frac{1-e^{-\tau}}{\tau}.
\end{equation}
Second, in the case of a static, spherically symmetric and homogeneous medium
the escape probability is (\citealt{osterbrock:book},
Appendix~2)\footnote{Some authors define the escape probability in
  terms of the optical radius $\tau_r$. This paper uses the optical
  diameter $\tau_d = 2\tau_r$ so that the comparison between
  geometries is more direct.}
\begin{equation}
\label{beta_sphere}
\beta_{\mathrm{sphere}} = \frac{1.5}{\tau}\left[1 - \frac{2}{\tau^2} + \left(\frac{2}{\tau} + \frac{2}{\tau^2}\right)e^{-\tau} \right].
\end{equation}

\begin{figure}[tb]
\centering
\resizebox{\hsize}{!}{\includegraphics[angle=-90]{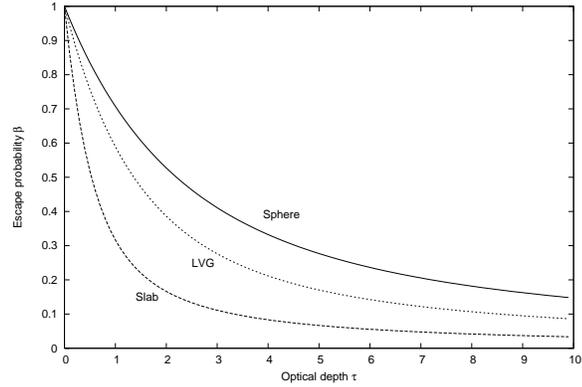}}  
\caption{Escape probability $\beta$ as a function of optical depth $\tau$ for
  three different geometries: uniform sphere (solid line), expanding sphere
  (dotted line) and plane-parallel slab (dashed line).}
\label{fig:beta}
\end{figure}

Third, for a plane-parallel `slab' geometry, applicable for instance to shocks,
\begin{equation}
\beta_{\mathrm{slab}} =  \frac{1-e^{-3\tau}}{3\tau}
\end{equation}
is derived \citep{dejong:collapse}. Figure~\ref{fig:beta} plots the behaviour of
$\beta$ as a function of $\tau$ for these three cases; \new{for more detailed
  comparisons see \citet{stutzki:winnewisser} and \cite{ossenkopf:radtrans}. }
Users of our program can select either expression for their calculations.  The
on-line version of the program uses the formula for the uniform sphere,
Eq.~(\ref{beta_sphere}).

\section{The program}
\label{s:radex} 
\texttt{RADEX} is a non-LTE radiative transfer code, written originally by
J.~H.~Black, that uses the escape probability formulation assuming an isothermal
and homogeneous medium without large-scale velocity fields. With the current
increase of observational possibilities in mind, we have developed a version of
this program which is suitable for public use. A guide for using the code in
practice is provided in Appendix~\ref{s:in-out} and
on-line\footnote{\texttt{http://www.sron.rug.nl/$\sim$vdtak/radex/ \\ index.shtml}};
Appendix~\ref{s:coding} describes the \old{steps from a private-use to a
  public-domain code} \new{adopted coding style}. 
This section focuses on the implementation of the formalism of \S~\ref{s:formal}
in the program.

\subsection{Basic capabilities}
\label{ss:capab}

For a homogeneous medium with no global velocity field, the optical depth at
line centre can be expressed using Eqs.~(\ref{sourcefuncdef}, \ref{localem},
\ref{sourcefunc}, \ref{lineprofile}), as
\begin{equation}
\tau = \frac{c^3}{8\pi\nu_{ul}^3}\frac{A_{ul}N_{\rm mol}}{1.064\Delta V}\left[ x_l\frac{g_u}{g_l}-x_u\right],
\end{equation}
where $N_{\rm mol}$ is the total column density, $\Delta V$ the full width at
half-maximum of the line profile in velocity units, and $x_i$ the fractional
population of level $i$.  The formalism is analogous to the LVG method, with the
global $n/(dV/dR)$ replaced by the local $N / \Delta V$, as in microturbulent codes
\citep{leung:liszt}.
The program iteratively solves the statistical equilibrium equations starting from
optically thin statistical equilibrium (\S~\ref{ss:calc}) for the initial level
populations.

\new{The program can handle up to seven collision partners simultaneously. In
  dense molecular clouds, \hh\ is the main collision partner for most species,
  but in some cases, separate cross sections may exist for collisions with the
  ortho and para forms of \hh, and electron collisions may be important for
  ionic species. In diffuse molecular clouds and PDRs, excitation by atomic H
  becomes important, particularly for fine structure lines, while for comets,
  \hho\ is the main collision partner. We refer to \citet{flower:collrates} for
  the basic theory of molecular collisions, and to \citet{dubernet:collrates} for an
  update of the latest results.}

The output of the program is the background-subtracted line intensity in
units 
%
%
%
of the
equivalent radiation temperature 
in the Rayleigh-Jeans limit. 
The background subtraction follows traditional cm- and mm-wave spectroscopic
observations where the differences between on-source and off-source
measurements are recorded, such that
\begin{equation}
T_{\mathrm R} = \frac{c^2}{2k\nu^2} \left(I^{\mathrm{em}}_{\nu} - I^{\mathrm{bg}}_{\nu}\right). 
\end{equation}
The radiation peak temperature $T_{\mathrm R}$ can be directly compared to the
observed antenna temperature corrected for the optical efficiency of the
telescope. However,
it should be emphasized that \texttt{RADEX} contains no information about the
geometry or length scale and that it is assumed that the source \old{is
  spatially resolved} \new{fills the antenna beam}. If the source is expected to
be smaller than the observational beam, computed line fluxes must be corrected
before comparing to observed fluxes.

In other types of observations, the continuum may not be subtracted from the
data.  In \smm\ and THz observations, for example with ESA's future
\textit{Herschel} space observatory, the dust continuum of many sources will be
much stronger than any instrumental error, and baseline subtraction may not be
needed.  The same is true for interferometer data, where the instrumental
passband is well characterized.  \old{To compare intensities from such
  observations with results from our program, the continuum would have to be
  subtracted first.}



\subsection{Background radiation field}
\label{ss:bg}

The average Galactic background (interstellar radiation field, ISRF) adopted in
\texttt{RADEX} consists of several components.  The main contribution is the
cosmic microwave background (CMB) whose absolute temperature is taken to be
$T_{\rm CBR}$ = 2.725$\pm$0.001\,K based on the full COBE data set as analyzed
by \citet{fixsen:cmb}.  This model of the microwave background represents the
broadband continuum only and does not include the strong emission lines,
several of which contain significant power in the far-infrared and \smm\ part of
the spectrum (see, e.g., \citealt{fixsen:cobe}).
\old{The other contributions to the \texttt{RADEX} background model are as follows.}
The ultraviolet/visible/near-infrared part of the spectrum is based on the model
of average Galactic starlight in the solar neighborhood of \citet{mathis:isrf}.
The far-infrared and \smm\ part of the spectrum is based on the
single-temperature fit to the Galactic thermal dust emission of
\citet{wright:isrf}.  At frequencies below 10\,\rcm\ (30\,GHz), there is a
background contribution from non-thermal radiation in the Galaxy.
A tabulation of this spectrum in ASCII format is available on-line\footnote{\tt
  http://www.oso.chalmers.se/$\sim$jblack/RESEARCH/ \\ isrf.dat}, and a graphical
representation is shown in \citet{black:isrf}.

One subtle aspect of the calculation is the distinction between the background
seen by the observer and the background seen by the molecules. The continuum
contribution to the rate equations may be composed of (1) an external component
which arises outside the emitting region and (2) an internal continuum that
arises within the emitting region. \old{A combination of the two is also possible.} 
The CMB and ISRF are examples of external continuum components; dust emission
from the line-emitting region is an example of an internal continuum. \new{While
an external continuum always fills the entire sky, an internal continuum may
only fill a fraction of it, for example in the case of a circumstellar disk.}

With this distinction in mind, the internal intensity becomes
\begin{equation}
  \label{eq:bgcomps}
J_{\nu}^{\rm int} = 
    \beta [B_{\nu}(T_{\rm CBR}) + \eta I_{\nu}^{\rm user}]
   + (1 - \beta) [B_{\nu}(T_{\rm ex}) + \theta (1-\eta) I_{\nu}^{\rm user}]  
\end{equation}
where $I_{\nu}^{\rm user}$ is the continuous spectrum defined by the user.
The factor $\eta$, 
is the fraction of local continuum which arises outside the line emitting region,
\new{and the factor $\theta$ is the fraction of local sky filled by the internal continuum.}

\subsection{Chemical formation and destruction rates}
\label{sec:chem}

The equations of statistical equilibrium (\ref{SE}) include source and sink
terms. By default, {\tt RADEX} sets the destruction rates equal to the same
small value, ${\cal D}_i \equiv {\cal D} = 10^{-15}$\,s$^{-1}$, appropriate for
cosmic-ray ionization plus cosmic-ray induced photodissociation
\citep{prasad:tarafdar,gredel:photodestruction}. The corresponding formation
rates are
\begin{equation}
  \label{eq:frate}
 {\cal F}_i = 10^{-24} n_{\rm total} g_i \exp(-E_i/kT_{\rm form}) /
Q(T_{\rm form}) 
\end{equation}
where $n_{\rm total}$ is the sum of the densities of all collision partners,
$T_{\rm form}$ is a formation temperature (default value 300\,K), and
\begin{equation}
  \label{eq:qpart}
 Q(T) = \sum_{i=1}^{N} g_i \exp(-E_i/kT_{\rm form})
\end{equation}
is the partition function. These assumptions imply a nominal 
fractional abundance of every molecule
\begin{equation}
  \label{eq:xdeft}
 {{n_{\rm mol}}\over n_{\rm total}} = {{\sum_i {\cal F}_i}\over
{n_{\rm total} {\cal D}}} = 10^{-9}\;\;\;. 
\end{equation}
The value of the nominal abundance is inconsequential because the results in
{\tt RADEX} depend on $N_{\rm mol}/\Delta V$, but not on the fractional
abundance. For most molecules currently in the associated database (LAMDA) and
for the most commonly encountered interstellar conditions, these choices will
not affect the observable excitation. The formation and destruction rates are
computed in a subroutine that can be modified by the user to provide a more
realistic description of chemical processes. 
For example, users may treat the combined ortho/para forms of molecules by
introducing a realistic $T_{\rm form}$, especially in cases where no $o/p$
interchange processes is likely to be effective.
\new{Other cases of potential interest include the photodissociation of large
  molecules into smaller molecules, or the evaporation of icy grain mantles into
the gas phase. Our formulation in terms of a volume rate of formation is chosen
to be independent of the details of the formation process.}
In general, formation and destruction processes are important for molecules that
have very long-lived (metastable) excited states and for molecules that are
expected to have very short chemical lifetimes.

\subsection{Calculation}
\label{ss:calc}

The input parameters of \texttt{RADEX} and its output are described in Appendix~\ref{s:in-out}.
Calculations with \texttt{RADEX} proceed as follows. A first guess of the
populations of the molecular energy levels is produced by solving statistical
equilibrium in the optically thin case. The only radiation taken into account is
the unshielded background radiation field; internally produced radiation is not
yet available. The solution for the level populations allows calculation of the
optical depths of all the lines, which are then used to re-calculate the
molecular excitation. The new calculation treats the background radiation in the
same manner as the internally produced radiation. The program iteratively finds
a consistent solution for the level populations and the radiation field. When
the optical depths of the lines with $\tau>10^{-2}$ are stable from one
iteration to the next to a given tolerance (default 10$^{-6}$), the program
writes output and stops. 

\begin{figure}
\centerline{\includegraphics[width=8cm]{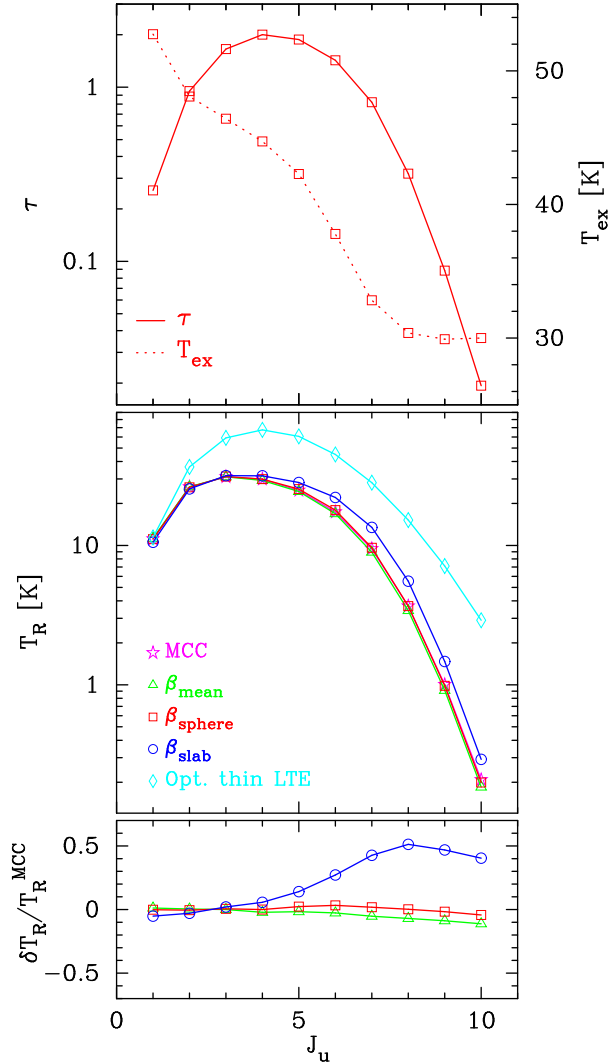}}  
\caption{Comparison of the predicted line strengths for the 10
lowest rotational transitions of CO for a homogeneous isothermal
sphere, with $n_{\mathrm{H_2}}=10^5$\,cm$^{-3}$ and
$T_{\mathrm{kin}}=50$\,K, using different methods. {\em Upper panel:}
The total optical depth through the sphere at line centre $\tau$
and excitation temperature $T_{\mathrm{ex}}$ as a function of the
upper rotational level $J$ involved in the transition.  {\em Middle
panel:} The radiation temperature $T_{\mathrm R}$ obtained for each
transition using \texttt{RADEX} with different prescriptions of the escape
probability $\beta$ and compared with the result from the Monte-Carlo
code (MCC) of \citet{schoeier:thesis}. Also shown are the results for
optically thin emission in LTE.  {\em Lower panel:} $T_{\mathrm R}$
obtained from \texttt{RADEX} compared with the results from MCC, $\delta
T_{\mathrm R}=T_{\mathrm R}^{\mathrm{MCC}}-T_{\mathrm R}$.}
  \label{compare}
\end{figure}

\subsection{Results}
\label{ss:rdx-results}

There are several ways in which \texttt{RADEX} can be used to analyze molecular
line observations. 
\new{In most of these applications, the modeled quantity is the
  velocity-integrated line intensity, as the excitation is assumed to be
  independent of velocity. As a consequence, self-absorbed lines cannot be
  modeled satisfactorily with \texttt{RADEX}.}
In the simplest case, the temperature and density are known from other
observations and only the column density of the molecule under consideration
needs to be varied to get the best agreement with the observed line intensity.
If the H$_2$ column density is known from other observations, for example from
an optically thin CO isotopic line, the ratio of the two column densities gives
the molecular abundance, averaged over the source.
The \texttt{RADEX} distribution contains a Python script to automate this
procedure \new{which is further described in Appendix~\ref{s:python}}. 

Another often-used application of \texttt{RADEX} is to determine temperatures
and densities from the observed intensity ratios of lines of the same molecule.
If the abundance of the molecule is constant throughout the source, the ratios
should give source-averaged physical conditions independent of the specific
chemistry of the molecule.  Appendix~\ref{s:diagn-plots-molec} presents
illustrative plots of line ratios for commonly observed molecules and lines in
the optically thin case. For higher optical depths, the qualitative trends
remain the same but there \old{will be minor} \new{are} quantitative differences. \texttt{RADEX}
can readily be used to generate similar plots for moderately thick cases.
Again, Python scripts are made available to automate this procedure
\new{(Appendix~\ref{s:python})}.

To illustrate the use of \texttt{RADEX} on actual observations, we take the
observations of the HCO$^+$ 1--0 and 3--2 lines toward a relatively
simple source, the photon-dominated region IC~63 \citep{jansen:ic63}.
The observed 1--0/3--2 ratio corrected for beam dilution is 5.5$\pm$1.5. The
kinetic temperature of the source is constrained from CO observations to be
$\sim$50\,K. Fig.~\ref{fig:hco+_ratio} shows that the inferred density for this
ratio is then $\sim 5\times 10^4$ cm$^{-3}$, a value confirmed by other line
ratios, e.g., CS 2--1/3--2. The inferred column density from the absolute
intensities is $8\times 10^{12}$ cm$^{-2}$, which, together with the overall
H$_2$ column density of $5\times 10^{21}$ cm$^{-2}$, gives an HCO$^+$ abundance
with respect to H$_2$ of $1.6\times 10^{-9}$.

A slightly more complicated situation arises for the Orion Bar PDR
\citep{hogerheijde:orion-bar}.  For this source, both HCO$^+$ 1--0, 3--2 and
4--3 lines have been observed. The 1--0/3--2 ratio gives an order of magnitude
lower density than the 3--2/4--3 ratio. This difference in density automatically
translates into an order of magnitude uncertainty in the inferred HCO$^+$ column
density, and thus the HCO$^+$ abundance, since column density tracers such as CO
are usually much less sensitive to density. \old{The proper} \new{One possible} solution is a clumpy PDR
model in which the 1--0 line is mostly produced in the low-density interclump
gas containing 90\% of the material and the 4--3 line in the high-density
clumps. Within this clumpy model, a single column density fits all three lines
and an accurate abundance can be derived.
\new{Note that this technique of adding the results of two models is only applicable at
  low optical depths.}

\new{If only two lines of a molecule have been observed, the line ratio can be
  used as indicator of temperature or density, depending on molecule and
  transition (Appendix~\ref{s:diagn-plots-molec}).
A single line ratio is never enough to constrain both temperature and density, though.
For multi-line observations, a comparison of data and models in terms of
$\chi^2$ is preferred. See for example \citet{vdtak:massive},
\citet{schoeier:16293}, and \citet{leurini:ch3oh} for details of such
calculations.}

\begin{figure}
\centerline{\includegraphics[width=7cm,angle=0]{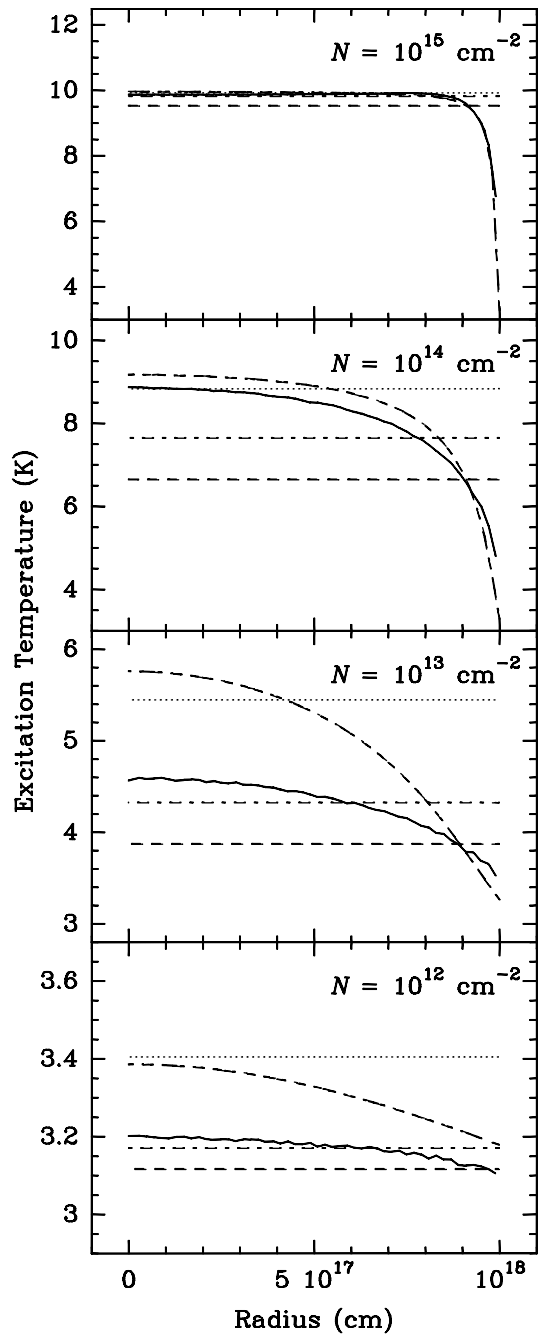}}
\caption{Excitation temperature of the \hcop\ 1--0 transition as a function of
  radius for the model \new{cloud} of \S~\ref{sss:mc-hcop}, calculated with \texttt{RADEX}
  assuming static spherical, expanding spherical, and slab geometry
  (\textit{dashed / dash-dotted / dotted lines}), with a multi-zone escape
  probability program (\textit{long/short dashes}) and with a Monte Carlo code
  (\textit{solid lines}). The panels are for different column densities,
    hence optical depths. Note the different vertical scales.}
\label{fig:radex_hst_txc}
\end{figure}

\subsection{Limitations of the program}
\label{ss:limits}

The current version of the program does not include a contribution from
continuous (dust or free-free) opacity to the escape probability, as for example
in \citet{takahashi:h2o}. Continuum radiation from dust is generally negligible
at long wavelengths ($\gtsim$1\,mm) but becomes important for regions with very
high column densities (such as protoplanetary disks) and at far-infrared and
shorter wavelengths ($\ltsim$100\,\mic). Free-free radiation may become
important for the calculation of atomic fine structure lines from \hii\ regions;
other programs such as \texttt{CLOUDY} \citep{ferland:cloudy}\footnote{\tt
  http://www.nublado.org/} may be more suitable for this purpose.
\new{The absence of continuous opacity limits the applicability of the program
  particularly in situations where infrared pumping is important, either
  directly through rotational transitions or via vibrational transitions
  \citep{carroll:pumping,hauschilt:cs10-9}.}

Another limitation of the program is that only one molecule is treated at a
time, so that the effects of line overlap are not taken into account. Such
overlaps may occur both at radio and at infrared wavelengths (e.g.,
\citealt{exposito:ir-overlap}). In special cases, overlap between lines of the
\textit{same} molecule may influence their excitation, for example the hyperfine
components of HCN or \nnhp\ \citep{daniel:hyperfine}.

For certain molecules \old{(particularly with complicated energy level structures)}
under certain physical conditions (especially low density and/or strong
radiation field), population inversions occur, which cause negative optical
depth and hence nonlinear amplification of the incoming radiation
\citep{elitzur:masers}. This phenomenon, known as `maser' action, requires
non-local treatment of the radiative transfer, in particular a fine sampling of
directions, for which \texttt{RADEX} is not set up. \new{Generally, the escape
  probability approximation is justified until the masers 
  saturate, which occurs at $\tau \approx -1$.} \old{If optical depths are
$\ltsim -$0.1, users of \texttt{RADEX} should check their results carefully.}
In practice, the computed intensities of lines with $\tau \ltsim -0.1$ are not
as accurate as those of other lines, and the intensities of lines with $\tau
\ltsim -1$ should be disregarded altogether. If many lines have negative optical
depths, the intensities of non-maser lines may also be affected. While
specialized programs should be used to calculate the intensities of maser lines
\citep[e.g.,][]{spaans:oh-masers,gray:maser,yates:maser}, \texttt{RADEX} may
\old{very} well be used to predict which lines of a molecule may display maser
action under certain physical conditions.
\new{Note however that lines may be masers even if $\tau > 0$ according to
  \texttt{RADEX}, for example `Class II' \meth\ masers which are pumped by
  infrared radiation \citep{leurini:ch3oh}.}

\section{Comparison with other methods}
\label{s:compare}

This section shows a comparison of \texttt{RADEX} with other programs, first for
the case of constant physical conditions (\S~\ref{ss:mc-rdx}) and second for
variable conditions (\S~\ref{ss:rd-rdx}). Comparison is with the analytical
rotation diagram method and with Monte Carlo methods, which have been
benchmarked to high accuracy, both for the case of \hcop\
\citep{zadelhoff:benchmark}\footnote{\tt
  http://www.strw.leidenuniv.nl/astrochem/ \\ radtrans/} and of \hho\
\citep{vdtak:h2o-benchmark}\footnote{\tt
  http://www.sron.rug.nl/$\sim$vdtak/H2O/}.
Throughout this section, molecular data have been taken from the LAMDA database
\citep{schoeier:lamda}. 

\subsection{Homogeneous models}
\label{ss:mc-rdx}

\subsubsection{The case of CO}
\label{sss:mc-co}
To test the \texttt{RADEX} code, we have compared its output both to an
optically thin LTE analysis (rotation diagram method) and a full radiative
transfer analysis using a \old{sophisticated non-LTE code based on the}
Monte-Carlo method \citep{schoeier:thesis}.  The test problem consists of a
spherically symmetric cloud with a constant density, $n$(\hh), of $1\times
10^5$\,cm$^{-3}$ within a radius of 100\,AU.  In this example only the CO
emission is treated using a fractional abundance of $1\times 10^{-4}$ relative
to H$_2$ yielding a \new{central} CO column density of $N_{\mathrm{CO}} = 3
\times 10^{16}$\,cm$^{-2}$ \new{and an average value of $N$=\pow{2}{16}\,\scm}.
The kinetic temperature is set to 50\,K, \new{the background temperature to
  2.73\,K, and the line width to \dv=1.0\,\kms}.

Fig.~\ref{compare} presents the results of the calculations for the ten lowest
rotational transitions. The excitation temperatures of the lines vary from being
close to thermalized for transitions involving low $J$-levels, to sub-thermally
excited for the higher-lying lines. The optical depth in the lines is moderate
($\sim 1-2$) to low. It is seen that the expressions of the escape probability
for the uniform sphere and the expanding sphere give almost identical solutions
which are close to that obtained from the full radiative transfer (MCC in
Fig.~\ref{compare}). The slab geometry gives slightly higher intensities, in
particular for high-lying lines. The optically thin approximation, where the gas
is assumed to be in LTE at 50\,K, produces much larger discrepancies, up to a
factor of $\sim 2$, and only gives the correct intensity for the $J=1\rightarrow
0$ line, where the LTE conditions are met.

\subsubsection{The case of  \hcop}
\label{sss:mc-hcop}

To further verify the performance of the \texttt{\texttt{RADEX}} program, we
have compared its results to that of another program that does not use the local
approximation: the Monte Carlo program \texttt{RATRAN} \citep{hogerheijde:montecarlo}. 
We also compare the results to those from the multi-zone escape probability
program by \citet{poelman:escprob}.
The test case is a cloud with $n$(\hh) = \pow{1}{4}\,\ccm, \tkin=10\,K, $T_{\rm
  bg}$=2.73\,K, and a line width of \dv=1.0\,\kms, equivalent to $b_D$=0.6\,\kms. 
The pure rotational emission spectrum of \hcop\ was calculated for column
densities of $10^{12}$, $10^{13}$, $10^{14}$ and $10^{15}$\,\scm, which for
\texttt{\texttt{RADEX}} were given directly as input parameters. For the
multi-zone programs, a cloud radius of $10^{18}$\,cm was specified along with
abundances of $10^{-10}$--$10^{-7}$, distributed over 50 cells.

\begin{figure}[tb]
\centering
\includegraphics[width=5cm,angle=-90]{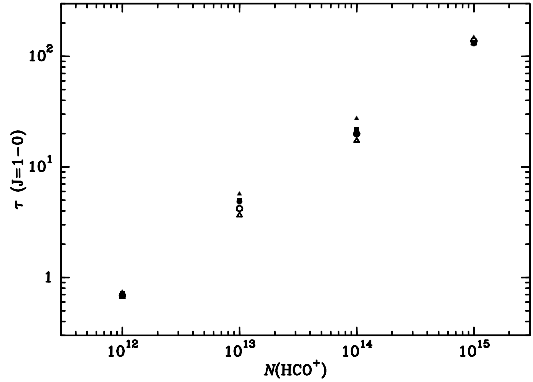}  
\caption{Optical depth of the \hcop\ 1--0 transition for the model of \S~\ref{sss:mc-hcop}, calculated with
  \texttt{RADEX} assuming static spherical, expanding spherical, and slab geometry
  (filled triangles/filled squares/open triangles), with the multi-zone escape
  probability code (open circles) and with the Monte Carlo code (filled circles).} 
\label{fig:radex_hst_tau}
\end{figure}

Figure~\ref{fig:radex_hst_txc} shows the calculated excitation temperature of
the \hcop\ 1--0 transition as a function of radius for these physical
conditions. For $N$(\hcop)$\ltsim$10$^{12}$\,\scm, the excitation is independent
of radius and the calculations for the various geometries agree to $\approx$10\%.
The dependence of the excitation on radius and on geometry increases with
increasing column density, and for $N$(\hcop)$\gtsim$10$^{15}$\,\scm, the
curvature of the \txc\ distribution becomes too large to ignore. The
corresponding line optical depth is $\approx$100, with $\approx$20\% spread between the
various estimates (Fig.~\ref{fig:radex_hst_tau}). The curvature arises because
at the cloud center, photon trapping thermalizes the excitation, while at the
edge, the emission can escape the cloud \citep{bernes:montecarlo}. 
We do not recommend to use \texttt{RADEX} at line optical depths
$\gtsim$100, because the calculated excitation temperature may not be
representative of the emitting region. However, even if some lines are highly
optically thick, \texttt{RADEX} may well be used to analyze other lines
which are optically thin. For example for \hho, the ground state lines often
have $\tau \sim 1000$, but \texttt{RADEX} is well capable of computing
intensities for higher-lying transitions which are not as optically thick.

At low optical depth, variations in \txc\ translate directly into changes in
emergent line intensity. 
Thus, differences as large as 20\% in \new{calculated line flux} can arise depending on the choice of
escape probability description, even for moderately thick cases.
At high optical depth, the direct connection between \txc\ and \new{line flux} is lost
because of the dependence on the adopted velocity field. The assumption in the
program that the optical depth is independent of velocity breaks down in this
case. \new{In this limit, the peak line temperature $T_R$ gives the value of
  \txc\ at the $\tau = 1$ surface of the cloud in this specific transition.}

The results shown in this section do not translate easily to other \hcop\ lines such as $J$=3$\to$2,
because the excitation is governed by several competing effects. The optical
depth of the $J$=3$\to$2 line may be higher or lower than that of the
$J$=1$\to$0 line, depending on temperature and density. Observers are encouraged
to use \texttt{RADEX} to study the excitation of their lines as a function of
these parameters, and also consider geometric variations.

\subsection{Observations of a Young Stellar Object}
\label{ss:rd-rdx}

\begin{figure}
  \centering
  \resizebox{\hsize}{!}{\includegraphics[width=8cm,angle=-90]{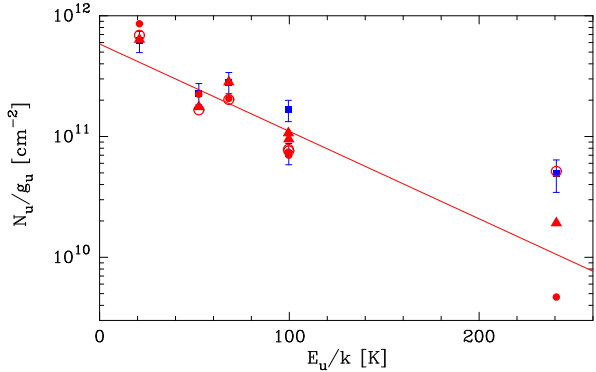}} 
  \caption{Line strengths of p-H$_2$CO observed towards the embedded low-mass
    protostar IRAS 16293--2422 (squares with error bars), modeled assuming LTE (solid
    line), using \texttt{RADEX} (triangles), and using a Monte Carlo program
    assuming a constant abundance (solid circles) and an abundance varying with
    radius (open circles).}
  \label{fig:rd-rdx}
\end{figure}

\begin{figure}
  \centering
  \centerline{\includegraphics[width=8cm]{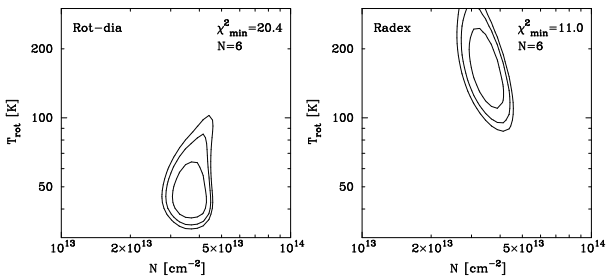}} 
  \caption{Distributions of the $\chi^2$ parameter corresponding to
    the models in Figure~\ref{fig:rd-rdx}. The \texttt{RADEX} results are for
    $n$(\hh) = 10$^6$\,\ccm\ as found by \cite{vdishoeck:16293}.}
  \label{fig:chisq}
\end{figure}

To compare a typical \texttt{RADEX} analysis with other methods \new{for a
  situation which varying physical conditions}, we choose a
molecule for which many lines can be observed: para-formaldehyde, p-\hhco. 
Figure~\ref{fig:rd-rdx} shows observations of p-\hhco\ \smm\ emission lines
originating from a wide range of energy levels toward the low-mass protostar
IRAS 16293--2422 by \citet{vdishoeck:16293}. The data are analyzed using
\old{two} \new{three}
methods: assuming LTE (with a rotation diagram), assuming SE (using
\texttt{RADEX}), \new{and using a Monte Carlo program}.
The free parameters for the LTE fit are the excitation temperature \txc\ and the
column density $N$(p-\hh CO). For the non-LTE fit, the free parameters are
kinetic temperature \tkin, \hh\ density $n$(\hh), and $N$(p-\hhco). 
\old{The observables are the intensities and the widths of the lines.}

Figure~\ref{fig:chisq} shows the distributions of the $\chi^2$ parameter for
\old{both} \new{the LTE and SE} fits, calculated in the standard way (see, e.g., \citealt{vdtak:massive};
\citealt{schoeier:16293}) assuming a 20\% uncertainty for all observed points
except the line with the highest upper level energy where a 30\% uncertainty was
used. As seen from the figure, the non-LTE method gives a better fit to the
data, as quantified by the lower minimum $\chi^2$ value. This result is not
necessarily surprising given that more free parameters are available.  A more
important difference is that the estimates of temperature and column density
between the two methods are substantially different, in particular the
temperature \new{(50 vs 150 K)}. Since the non-LTE method involves fewer assumptions about the
physical state of the cloud, its results are to be preferred.

These results illustrate that rotation diagrams may give misleading results when
determining physical properties of interstellar gas clouds
(cf.\ \citealt{johnstone:astrochem} for the case of \meth).
Figure~\ref{fig:rd-rdx} also demonstrates that temperatures and column densities
derived from rotation diagrams tend to depend on which lines happen to have been
observed \new{(cf.\ the \hcop\ case in \S~\ref{sss:mc-hcop})}.
From other data, IRAS 16293--2422 is actually known to have a gradient in
temperature and density throughout its envelope, which cannot be modelled
properly with either technique. For such situation, a full Monte Carlo radiative
transfer method is needed in which both the physical conditions and the
abundances can vary with radius (Fig.~\ref{fig:rd-rdx}, circles). Nevertheless,
the column densities and abundances inferred with \texttt{RADEX} using the
physical conditions inferred from the line ratios differ by only a factor of a
few from those found with the more sophisticated analysis, at least for the
particular zone of the source to which those conditions apply
\citep{schoeier:16293}.

\section{Conclusions}
\label{s:concl}

We have presented a computer program to analyze spectral line observations at
radio and infrared wavelengths, based on the escape probability \old{method} \new{approximation}.  The
program can be used for any molecule for which collisional data exist; such
input data are available in the required format from the LAMDA database. 
The program can be used for optical depths from $\approx$--0.1 to $\approx$100.

The limited number of free parameters makes \texttt{RADEX} very useful to
rapidly analyze large datasets. As an example, observed line intensity ratios
may be compared with the plots in Appendix~\ref{s:diagn-plots-molec} to estimate
density and kinetic temperature. Ratios of other lines and other molecules may
be easily computed using the Python scripts included in the \texttt{RADEX}
distribution.
The program may also be used to create synthetic spectra.  This capability will
be important to model the THz line surveys from the HIFI instrument onboard the
\textit{Herschel} space observatory.

In the \old{near} future, we plan to incorporate a multi-zone escape probability
formalism \citep{poelman:escprob,elitzur:multi} which will \new{enable models
  with varying physical conditions and also} improve the performance of the
program at high optical depths. To speed up convergence at high optical depths,
the calculation may also start from LTE conditions rather than from optically
thin statistical equilibrium.
Robust convergence may be achieved by starting from either initial condition and
requiring the two answers to be equal.
\old{In the more distant future, we plan to add the possibility to consider molecules
for which calculated collisional rate coefficients are not available.  With this
capability, }
For the modeling of crowded spectra, the
effects of line overlap will also need to be considered, for instance in the `all
or nothing' approach \citep{cesaroni:overlap}. Such spectra will be routinely
observed with the superb resolution and sensitivity of ALMA.

Our program is free for anybody to use for \old{private purposes} \new{science}, provided that
appropriate reference is made to this paper. For any other purpose such as to
incorporate the program into other packages which may be distributed to the
public, prior agreement with the authors is needed.

\begin{acknowledgements}
  The authors wish to thank Huib Jan van Langevelde for his efforts in
  documenting \texttt{RADEX}, and Erik Deul for computing support at Leiden
  Observatory. JHB and FLS acknowledge the Swedish Research Council for
  financial support. \new{FvdT and EvD thank the Netherlands Organization for
    Scientific Research (NWO) and the Netherlands Research School for Astronomy
    (NOVA). Finally we thank Volker Ossenkopf, Marco Spaans, and an
    anonymous referee for helpful comments on the manuscript.}
\end{acknowledgements}

\bibliographystyle{aa}
\bibliography{radex}

\Online

\appendix

\section{Program input and output}
\label{s:in-out}

\subsection{Program input}
\label{ss:input}

The input parameters to \texttt{RADEX} are the following:

\begin{enumerate}
\item The name of the molecular data file to be used.
\item The name of the file to write the output to.
\item The frequency range for the output file [GHz]. All transitions from the
  molecular data file are always taken into account in the calculation, but
  often it is practical to write only a limited set of lines to the output.
\item The kinetic temperature of the cloud [K].
\item The number of collision partners to be used. Most users will want \hh\ as
  only collision partner, but in more specialized cases, additional collisions
  with H or electrons may for instance play a role. See the molecular datafiles
  for details. For some species (CO, atoms) separate collision data for ortho
  and para \hh\ exist; the program then uses the thermal ortho/para ratio unless
  the user specifies otherwise.
\item The name (case-insensitive) and the density [\ccm] of each collision
  partner. Possibilities are \hh, p-\hh, o-\hh, electrons, atomic H, He, and H$^+$.
\item The temperature of the background radiation field [K]. 
  \begin{itemize}
  \item If $>$0, a black body at this temperature is used. Most users will adopt
    the cosmic microwave background at $T_{\rm CMB}$=2.725(1+$z$)\,K for a galaxy
    at redshift $z$.
  \item If =0, the average interstellar radiation
  field (ISRF) is used, taken from \citet{black:isrf} with modifications
  described in \S~\ref{ss:bg}. This spectrum is not adjustable by a scale
  factor because it consists of several components that are not expected to
  scale linearly with respect to each other.  
  \item If $<$0, a user-defined radiation
  field is used, specified by values of frequency [\rcm], intensity [Jy
  nsr$^{-1}$], and dilution factor [dimensionless]. Spline interpolation
  \new{and extrapolation are} 
  applied to this table. The intensity need not be specified at all frequencies
  of the line list, but a warning message will appear if extrapolation (rather
  than interpolation) is required.
  \end{itemize}
\item The column density of the molecule [\scm].
\item The FWHM line width [\kms].
\end{enumerate}

\subsection{Program output}
\label{ss:output}

The output file written by \texttt{RADEX} first replicates the input parameters,
and then lists the following quantities for each spectral line within the
specified frequency range.

\begin{enumerate}
\item Quantum numbers, upper state energy [K], frequency [GHz], and wavelength
  [\mic]. These numbers are just copied from the molecular data file, which
  usually comes from the LAMDA database.  Frequencies from this database are
  generally of spectroscopic accuracy ($\ltsim$0.1\,MHz uncertainty), although
  for the most precise values to be used for observations, the original
  literature and current line catalogs such as CDMS
  \citep{mueller:cdms}\footnote{\tt http://cdms.de} should be consulted.
\item The excitation temperature [K] as defined in Eq.~(\ref{Boltzmann}). In
  general, different lines have different excitation temperatures. Lines are
  thermalized if \txc=\tkin; in LTE, all lines are thermalized.
\item The line optical depth, defined as the optical depth of the equivalent
  rectangular line shape ($\phi_\nu=1/\Delta\nu$).
\item The line intensity, defined as the Rayleigh-Jeans equivalent temperature
  $T_R$ [K].
\item The line flux, defined as the velocity-integrated intensity, both in units
  of K\,\kms\ (common in radio astronomy) and of erg\,\scm\,s$^{-1}$ (common in
  infrared astronomy). The line flux is calculated as 1.0645$T_R$\dv,
  where the factor 1.0645 = $\sqrt{\pi}/(2\sqrt{\ln2})$ converts the adopted
  rectangular line profile into a Gaussian profile with an FWHM of \dv. The
  integrated profile is useful to estimate the total emission in the line, but
  it has limited meaning at high optical depths, because the change of optical
  depth over the line profile is not taken into account. Proper modeling of
  optically thick lines requires programs that resolve the source both
  spectrally and spatially (see \S~\ref{sss:mc-hcop} for further discussion).
\end{enumerate}

Auxiliary output files can be generated, for example to display the adopted
continuum spectrum.

\section{Coding standards}
\label{s:coding}

The original version of \texttt{RADEX} was written in such a way as to minimize
the use of machine memory which was expensive until a decade ago. Nowadays,
clarity and easy maintenance are more important requirements, which is why the
source code has been re-written following the rules below. We hope that these
rules will be useful for the development of other `open source' astronomical
software. For further guidelines on scientific programming we recommend the
Software Carpentry\footnote{\tt http://www.swc.scipy.org/} on-line course.

\begin{enumerate}
\item All the action is in subroutines; the sole purpose of the main
  program is to show the structure of the program. The subroutines are grouped
  into several files for a better overview; compilation instructions for
  automated builds on a variety of platforms are in a Makefile.
\item The program text is interspersed with comments at a ratio of
  $\approx$1:1. In particular, each subroutine starts with a
  description of its contents, its input and output, where incoming
  calls come from, and which calls go out. Then the properties of each
  variable are described: contents, units and type.
\item Variables and subroutines have descriptive names with a length of 5--10
  letters. Names of integer variables start with the letters i..n; names of
  floating-point variables (always of double precision) with a..h or o..z. There
  are no specific namings for variables of character or logical type, as for
  example in \texttt{CLOUDY}. Names are always based on the English language.
  We do not use upper-, lower-, or camel-case to distinguish types of
  identifiers, as some programs do.
\item Loops are marked by indenting the program text. The loop variables are
  always called ilev, iline .., never just i.
\item Subroutines start with a check whether the input parameters have
  reasonable values. Such checks force soft landings if necessary, and avoid
  runtime errors.
\item Statements with calculations use spaces around the =, + and -- symbols, but
  not around others. Calculations that consist of multiple steps are split over
  as many program lines. Multiple assignments in a row are aligned at the =
  sign. 
\item The program text avoids ``magic numbers'' both in calculations and in
  definitions. Numbers that are often used such as Planck's constant are defined
  at one central place in the program. Similarly, often-used variables such as
  physical parameters are stored in shared memory rather than passed on via
  subroutine calls.
\end{enumerate}

\newpage

\section{Diagnostic plots of molecular line ratios}
\label{s:diagn-plots-molec}

We have used the \texttt{RADEX} program and the LAMDA database to calculate
line ratios of several commonly observed molecules for a range of
kinetic temperatures and \hh\ densities. The plots in this Appendix
may be used by observers to estimate physical conditions from their
data. 
Line ratios have the advantage of being less sensitive to
calibration errors than absolute line strengths, especially when
the two lines have been measured with the same telescope, receiver and
spectrometer. 

The calculations assume a column density of 10$^{12}$\,\scm, a
line width of 1.0\,\kms, and use a 2.73\,K blackbody as background
radiation field. Under these conditions, the lines are optically thin,
so that the line ratios do not depend on column density.
The calculations also assume that the emission in both lines fills the
telescope beams equally, which may be the case if the lines are close
in frequency.
However, lines are generally measured in beams of different sizes, and
the observations need to be corrected to account for this effect, if the
source is known to be compact.

\begin{figure}[p]
\centering
\resizebox{\hsize}{!}{\includegraphics[angle=0]{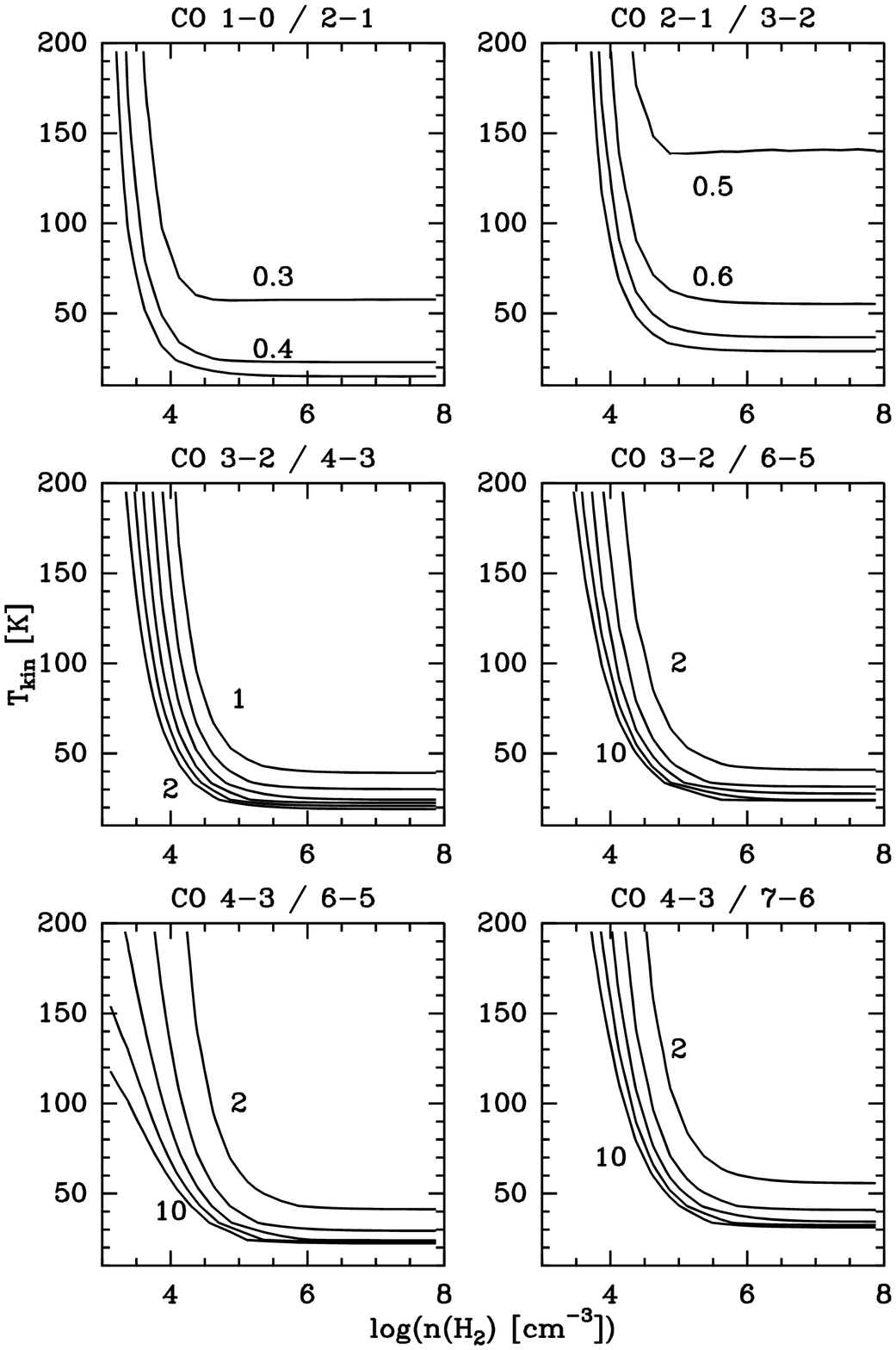}}  
\caption{Line ratios of CO \new{in the optically thin limit} as a function of kinetic temperature and
  \hh\ density. Contours are spaced linearly and some contours are
  labeled for easy identification.} 
\label{fig:co_ratio}
\end{figure}

Linear molecules such as CO (Fig.~\ref{fig:co_ratio}) are tracers of
density at low densities, when collisions compete with radiative
decay. At higher densities, the excitation becomes thermalized and the
line ratios are sensitive to temperature. For a given molecule, moving
up the $J-$ladder means probing higher temperatures and densities.
Note that for the column densities of typical dense interstellar
clouds, the CO lines are optically thick, and observations of \thc O
or even rarer isotopologues must be used to probe physical conditions.

\begin{figure}[p]
\centering
\resizebox{\hsize}{!}{\includegraphics[angle=0]{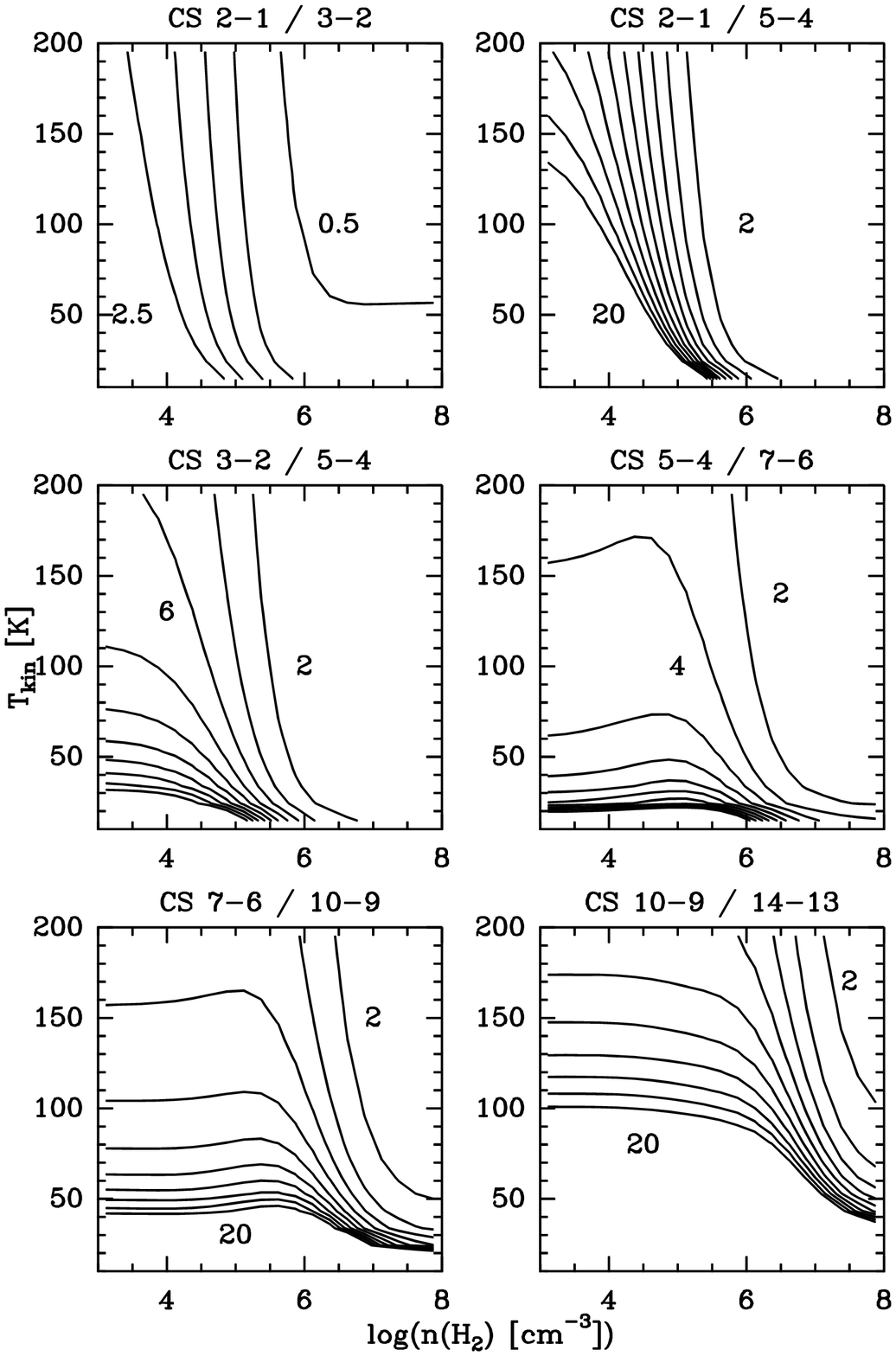}}  
\caption{Line ratios of CS \new{in the optically thin limit} as a function of kinetic temperature and
  \hh\ density. Contours are spaced linearly and some contours are
  labeled for easy identification.} 
\label{fig:cs_ratio}
\end{figure}

\begin{figure}[p]
\centering
\resizebox{\hsize}{!}{\includegraphics[angle=0]{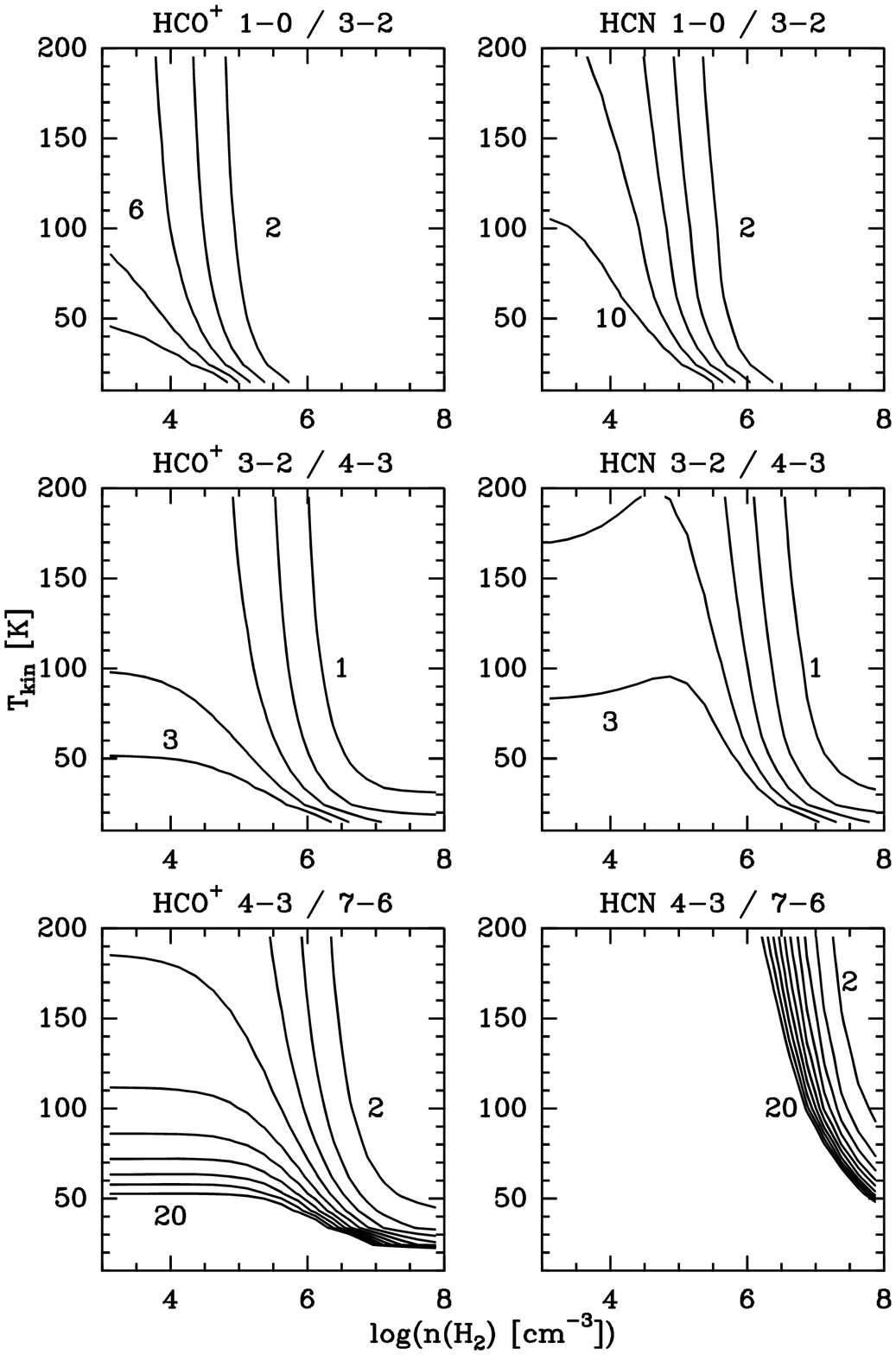}}  
\caption{Line ratios of \hcop\ and HCN \new{in the optically thin limit} as a function of kinetic temperature and
  \hh\ density. Contours are spaced linearly and some contours are
  labeled for easy identification.} 
\label{fig:hco+_ratio}
\end{figure}

The critical densities of molecular lines scale as $\mu^2\nu^3$, where $\mu$ is
the permanent dipole moment of the molecule and $\nu$ is the frequency of the
line. 
Indeed, the CS molecule (Fig.~\ref{fig:cs_ratio}) has a larger dipole moment
than CO, and its line ratios are mainly probes of the density.
The small frequency spacing between the lines of CS makes this molecule very
useful to probe density structure (e.g., \citealt{vdtak:massive}).
The \hcop\ and HCN molecules (Fig.~\ref{fig:hco+_ratio}) display
similar trends to CS, although their line spacing is not as small.

\begin{figure}[p]
\centering
\resizebox{\hsize}{!}{\includegraphics[angle=0]{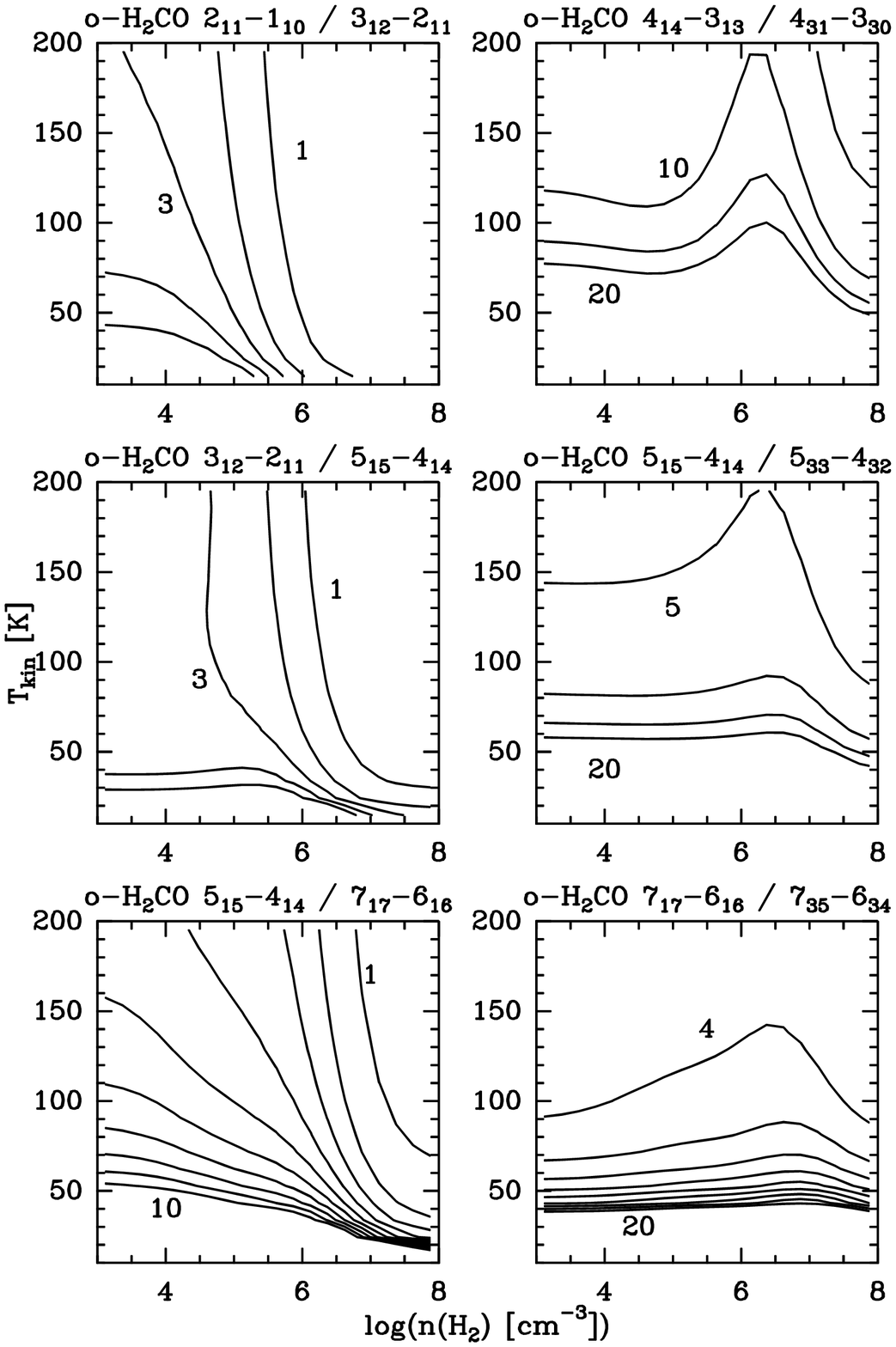}}  
\caption{Line ratios of o-\hh CO \new{in the optically thin limit} as a function of kinetic temperature and
  \hh\ density. Contours are spaced linearly and some contours are
  labeled for easy identification.} 
\label{fig:oh2co_ratio}
\end{figure}

\begin{figure}[p]
\centering
\resizebox{\hsize}{!}{\includegraphics[angle=0]{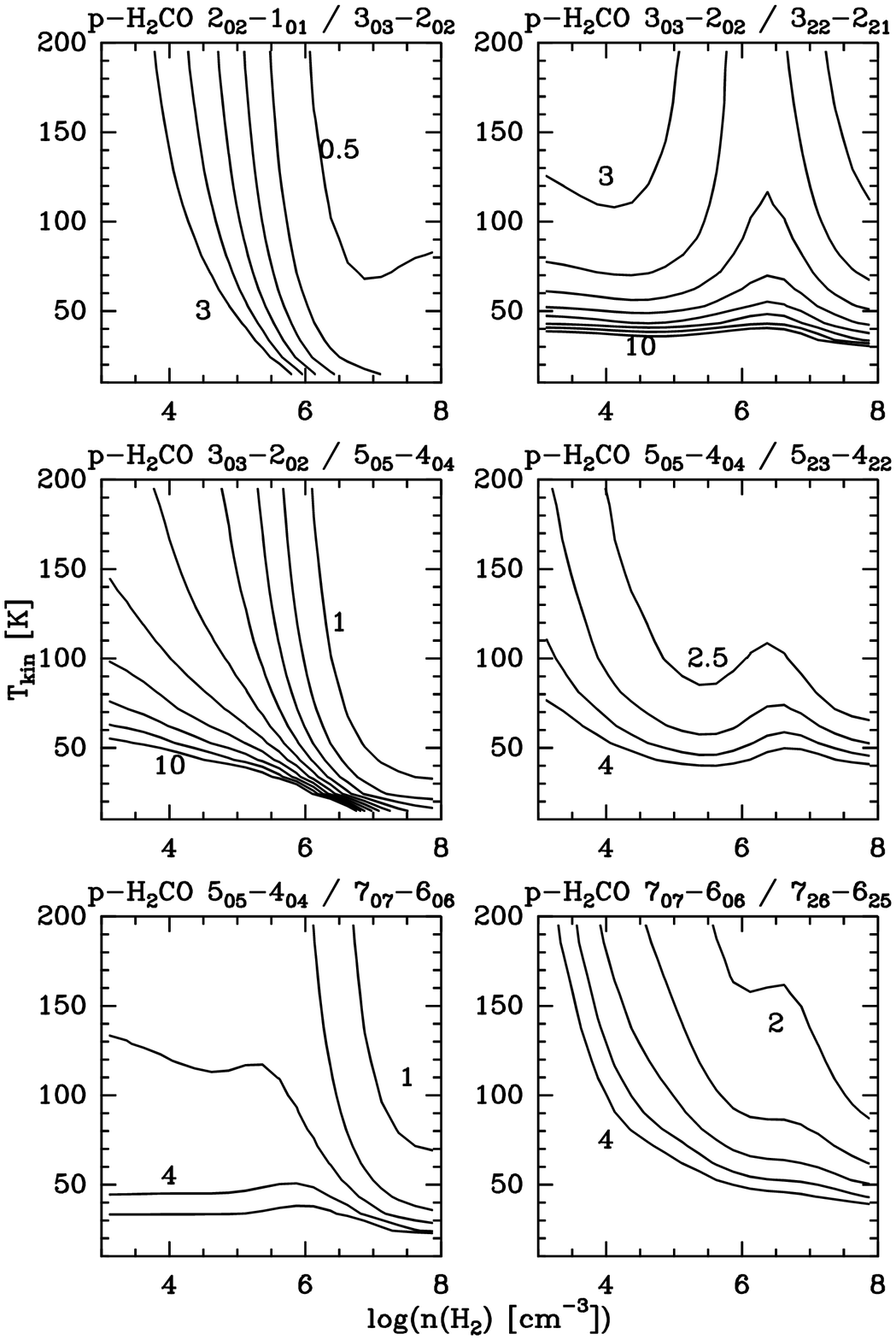}}  
\caption{Line ratios of p-\hh CO \new{in the optically thin limit} as a function of kinetic temperature and
  \hh\ density. Contours are spaced linearly and some contours are
  labeled for easy identification.} 
\label{fig:ph2co_ratio}
\end{figure}

Non-linear molecules such as \hhco\ (Figs.~\ref{fig:oh2co_ratio}
and~\ref{fig:ph2co_ratio}) have the advantage that both temperature and density
may be probed within the same frequency range. Ratios of lines from different
$J-$states tend to be density tracers (left panels), while ratios of lines from
the same $J-$state but different $K-$states are mostly temperature probes (right
panels).
The lines of \hhco\ are often quite strong, making this molecule a favourite
tracer of temperature and density \citep{mangum:h2co}.  Other asymmetric
molecules have also been used, such as \hhcs\ \citep{blake:16293} and \meth\
\citep{leurini:ch3oh} although abundance variations from source to source or
even within sources often complicate the interpretation
\citep{johnstone:astrochem}.

\section{The Python scripts}
\label{s:python}

The \radex\ distribution comes with two scripts, \texttt{radex\_line.py} and \texttt{radex\_grid.py},
to automate standard modeling procedures. The scripts are written in Python and
are run from the Unix shell command line after manual editing of parameters.

\bigskip

The first script, \texttt{radex\_line.py}, calculates the column density of a
molecule from an observed line intensity, given estimates of kinetic temperature
and \hh\ volume density. The input parameters are:

\begin{enumerate}
\item The kinetic temperature [K]
\item The number density of \hh\ molecules [\ccm]
\item The temperature of the background radiation field [K], usually 2.73 (CMB).
\item The name of the molecule (or molecular data file)
\item The frequency of the line [GHz]
\item The observed line intensity [K]
\item The observed line width [\kms]
\end{enumerate}

Furthermore, two numerical parameters have good default values but will need to
be changed occasionally:

\begin{enumerate}
\item The free spectral range around the line (default 10\%): this number must
  be smaller for molecules with many line close in frequency, such as \meth. 
  The program uses this parameter to find the observed line from the list of
  lines in the molecular model.
\item The required accuracy (default 10\%): The default corresponds to the
  calibration uncertainty of most telescopes. 
\end{enumerate}

The script iterates on column density until the observed and modeled line fluxes
agree to within the desired accuracy.  The best-fit column density is directly
written to the screen. The file \texttt{radex.out} gives details of the best-fit
model.

\bigskip

The second script, \texttt{radex\_grid.py}, runs a series of \radex\ models to
estimate the kinetic temperature and/or the volume density from an observed line
ratio. The user needs to set the following input parameters:

\begin{enumerate}
\item The grid boundaries: minimum and maximum kinetic temperature [K] and
  minimum and maximum \hh\ volume density [\ccm].
\item The temperature of the background radiation field [K], usually 2.73 (CMB).
\item The molecular column density [\scm]. For the illustrative plots in
  Appendix~\ref{s:diagn-plots-molec}, a low value ($N$=10$^{12}$\,\scm) was
  used, so that the line ratios are independent of column density (optically
  thin limit). However, in modeling specific observations, it is worth varying
  this parameter to assess the sensitivity of the line ratio to column density.
\item The observed line width [\kms], usually an average of the widths of the
  two lines.
\end{enumerate}

The numerical parameters which have good default values but will need to
be changed occasionally are:

\begin{enumerate}
\item The number of grid points along the temperature and density axes.
\item The free spectral range around the line (see above)
\end{enumerate}

The name of the molecule and a list of observed line ratios and names of
associated output files are given at the start of the main program. The script
produces a file \texttt{radex.out} which is a tabular listing of temperature,
log density, and line ratio. This results may be plotted with the user's
favourite plotting program.

\end{document}

%% file: defs.tex
\def\hii{H~{\sc II}}

\def\hh{H$_2$}

\def\thc{$^{13}$C}
\def\hho{H$_2$O}

\def\hcop{HCO$^+$}

\def\hhco{H$_2$CO}
\def\hhcs{H$_2$CS}
\def\meth{CH$_3$OH}

\def\hhhp{H$_3^+$}

\def\nnhp{N$_2$H$^+$}

\def\smm{(sub-)mil\-li\-me\-ter}
\def\pow#1#2{#1$\times$10$^{#2}$}

\def\gtsim{{_>\atop{^\sim}}}
\def\ltsim{{_<\atop{^\sim}}}

\def\tkin{$T_{\rm kin}$}
\def\dv{$\Delta$\textit{V}}
\def\kms{km~s$^{-1}$}

\def\txc{$T_{\rm ex}$}
\def\rcm{cm$^{-1}$}
\def\scm{cm$^{-2}$}
\def\ccm{cm$^{-3}$}

\def\mic{$\mu$m}

\def\aap{{A\&A}}

\def\apjl{{ApJ}}
\def\apjs{{ApJS}}



%% file: aa6820.bbl
\begin{thebibliography}{75}
\expandafter\ifx\csname natexlab\endcsname\relax\def\natexlab#1{#1}\fi

\bibitem[{{Bernes}(1979)}]{bernes:montecarlo}
{Bernes}, C. 1979, \aap, 73, 67

\bibitem[{{Black}(1994)}]{black:isrf}
{Black}, J.~H. 1994, in ASP Conf. Ser. 58: The First Symposium on the Infrared
  Cirrus and Diffuse Interstellar Clouds, ed. R.~M. {Cutri} \& W.~B. {Latter},
  355

\bibitem[{{Black}(1998)}]{black:faraday}
{Black}, J.~H. 1998, in Chemistry and Physics of Molecules and Grains in Space.
  Faraday Discussions No. 109, 257

\bibitem[{{Black}(2000)}]{black:korea}
{Black}, J.~H. 2000, in IAU Symposium 197 -- Astrochemistry: From Molecular
  Clouds to Planetary Systems, ed. Y.~C. {Minh} \& E.~F. {van Dishoeck}, 81

\bibitem[{{Black} \& {van Dishoeck}(1987)}]{black:h2}
{Black}, J.~H. \& {van Dishoeck}, E.~F. 1987, \apj, 322, 412

\bibitem[{{Blake} {et~al.}(1987){Blake}, {Sutton}, {Masson}, \&
  {Phillips}}]{blake:orion}
{Blake}, G.~A., {Sutton}, E.~C., {Masson}, C.~R., \& {Phillips}, T.~G. 1987,
  \apj, 315, 621

\bibitem[{{Blake} {et~al.}(1994){Blake}, {van Dishoek}, {Jansen}, {Groesbeck},
  \& {Mundy}}]{blake:16293}
{Blake}, G.~A., {van Dishoek}, E.~F., {Jansen}, D.~J., {Groesbeck}, T.~D., \&
  {Mundy}, L.~G. 1994, \apj, 428, 680

\bibitem[{{Burton} {et~al.}(1990){Burton}, {Hollenbach}, \&
  {Tielens}}]{burton:h2pumping}
{Burton}, M.~G., {Hollenbach}, D.~J., \& {Tielens}, A.~G.~G.~M. 1990, \apj,
  365, 620

\bibitem[{{Cannon}(1985)}]{cannon:book}
{Cannon}, C.~J. 1985, {The transfer of spectral line radiation} (Cambridge:
  University Press)

\bibitem[{{Carroll} \& {Goldsmith}(1981)}]{carroll:pumping}
{Carroll}, T.~J. \& {Goldsmith}, P.~F. 1981, \apj, 245, 891

\bibitem[{{Castor}(1970)}]{castor:escprob}
{Castor}, J.~I. 1970, \mnras, 149, 111

\bibitem[{{Cesaroni} \& {Walmsley}(1991)}]{cesaroni:overlap}
{Cesaroni}, R. \& {Walmsley}, C.~M. 1991, \aap, 241, 537

\bibitem[{{Comito} {et~al.}(2005){Comito}, {Schilke}, {Phillips}, {Lis},
  {Motte}, \& {Mehringer}}]{comito:survey}
{Comito}, C., {Schilke}, P., {Phillips}, T.~G., {et~al.} 2005, \apjs, 156, 127

\bibitem[{{Daniel} {et~al.}(2006){Daniel}, {Cernicharo}, \&
  {Dubernet}}]{daniel:hyperfine}
{Daniel}, F., {Cernicharo}, J., \& {Dubernet}, M.-L. 2006, \apj, 648, 461

\bibitem[{{De Jong} {et~al.}(1980){De Jong}, {Boland}, \&
  {Dalgarno}}]{dejong:lvg}
{De Jong}, T., {Boland}, W., \& {Dalgarno}, A. 1980, \aap, 91, 68

\bibitem[{{De Jong} {et~al.}(1975){De Jong}, {Dalgarno}, \&
  {Chu}}]{dejong:collapse}
{De Jong}, T., {Dalgarno}, A., \& {Chu}, S.-I. 1975, \apj, 199, 69

\bibitem[{{Doty} {et~al.}(2004){Doty}, {Sch{\"o}ier}, \& {van
  Dishoeck}}]{doty:16293}
{Doty}, S.~D., {Sch{\"o}ier}, F.~L., \& {van Dishoeck}, E.~F. 2004, \aap, 418,
  1021

\bibitem[{{Dubernet}(2005)}]{dubernet:collrates}
{Dubernet}, M.~L. 2005, in IAU Symposium, ed. D.~C. {Lis}, G.~A. {Blake}, \&
  E.~{Herbst}, 235

\bibitem[{{Elitzur}(1992)}]{elitzur:masers}
{Elitzur}, M. 1992, {Astronomical masers} (Kluwer Academic Publishers)

\bibitem[{{Elitzur} \& {Asensio Ramos}(2006)}]{elitzur:multi}
{Elitzur}, M. \& {Asensio Ramos}, A. 2006, \mnras, 365, 779

\bibitem[{{Evans} {et~al.}(2005){Evans}, {Lee}, {Rawlings}, \&
  {Choi}}]{evans:b335}
{Evans}, II, N.~J., {Lee}, J.-E., {Rawlings}, J.~M.~C., \& {Choi}, M. 2005,
  \apj, 626, 919

\bibitem[{{Exp{\'o}sito} {et~al.}(2006){Exp{\'o}sito}, {Ag{\'u}ndez},
  {Tercero}, {Pardo}, \& {Cernicharo}}]{exposito:ir-overlap}
{Exp{\'o}sito}, J.~P.~F., {Ag{\'u}ndez}, M., {Tercero}, B., {Pardo}, J.~R., \&
  {Cernicharo}, J. 2006, \apjl, 646, L127

\bibitem[{{Ferland}(2003)}]{ferland:cloudy}
{Ferland}, G.~J. 2003, \araa, 41, 517

\bibitem[{{Fixsen} {et~al.}(1999){Fixsen}, {Bennett}, \&
  {Mather}}]{fixsen:cobe}
{Fixsen}, D.~J., {Bennett}, C.~L., \& {Mather}, J.~C. 1999, \apj, 526, 207

\bibitem[{{Fixsen} \& {Mather}(2002)}]{fixsen:cmb}
{Fixsen}, D.~J. \& {Mather}, J.~C. 2002, \apj, 581, 817

\bibitem[{{Flower}(1989)}]{flower:collrates}
{Flower}, D.~R. 1989, Physics Reports, 174, 1

\bibitem[{{Fuente} {et~al.}(2000){Fuente}, {Black}, {Mart{\'{\i}}n-Pintado},
  {Rodr{\'{\i}}guez-Franco}, {Garc{\'{\i}}a-Burillo}, {Planesas}, \&
  {Lindholm}}]{fuente:co+}
{Fuente}, A., {Black}, J.~H., {Mart{\'{\i}}n-Pintado}, J., {et~al.} 2000,
  \apjl, 545, L113

\bibitem[{{Genzel}(1991)}]{genzel:crete1}
{Genzel}, R. 1991, in NATO ASIC Proc. 342: The Physics of Star Formation and
  Early Stellar Evolution, ed. C.~J. {Lada} \& N.~D. {Kylafis}, 155

\bibitem[{{Goicoechea} {et~al.}(2006){Goicoechea}, {Pety}, {Gerin}, {Teyssier},
  {Roueff}, {Hily-Blant}, \& {Baek}}]{goicoechea:horsehead}
{Goicoechea}, J.~R., {Pety}, J., {Gerin}, M., {et~al.} 2006, \aap, 456, 565

\bibitem[{{Goldreich} \& {Scoville}(1976)}]{goldreich:lvg}
{Goldreich}, P. \& {Scoville}, N. 1976, \apj, 205, 144

\bibitem[{{Goldsmith} \& {Langer}(1999)}]{goldsmith:population}
{Goldsmith}, P.~F. \& {Langer}, W.~D. 1999, \apj, 517, 209

\bibitem[{{Gray} \& {Field}(1995)}]{gray:maser}
{Gray}, M.~D. \& {Field}, D. 1995, \aap, 298, 243

\bibitem[{{Gredel} {et~al.}(1989){Gredel}, {Lepp}, {Dalgarno}, \&
  {Herbst}}]{gredel:photodestruction}
{Gredel}, R., {Lepp}, S., {Dalgarno}, A., \& {Herbst}, E. 1989, \apj, 347, 289

\bibitem[{{Hauschildt} {et~al.}(1993){Hauschildt}, {G\"usten}, {Phillips},
  {Schilke}, {Serabyn}, \& {Walker}}]{hauschilt:cs10-9}
{Hauschildt}, H., {G\"usten}, R., {Phillips}, T.~G., {et~al.} 1993, \aap, 273,
  L23

\bibitem[{{Helmich} {et~al.}(1994){Helmich}, {Jansen}, {de Graauw},
  {Groesbeck}, \& {van Dishoeck}}]{helmich:w3}
{Helmich}, F.~P., {Jansen}, D.~J., {de Graauw}, T., {Groesbeck}, T.~D., \& {van
  Dishoeck}, E.~F. 1994, \aap, 283, 626

\bibitem[{{Helmich} \& {van Dishoeck}(1997)}]{helmich:survey}
{Helmich}, F.~P. \& {van Dishoeck}, E.~F. 1997, \aaps, 124, 205

\bibitem[{{Hogerheijde} {et~al.}(1995){Hogerheijde}, {Jansen}, \& {van
  Dishoeck}}]{hogerheijde:orion-bar}
{Hogerheijde}, M.~R., {Jansen}, D.~J., \& {van Dishoeck}, E.~F. 1995, \aap,
  294, 792

\bibitem[{{Hogerheijde} \& {van der Tak}(2000)}]{hogerheijde:montecarlo}
{Hogerheijde}, M.~R. \& {van der Tak}, F.~F.~S. 2000, \aap, 362, 697

\bibitem[{{Jakob} {et~al.}(2007){Jakob}, {Kramer}, {Simon}, {Schneider},
  {Ossenkopf}, {Bontemps}, {Graf}, \& {Stutzki}}]{jakob:dr21}
{Jakob}, H., {Kramer}, C., {Simon}, R., {et~al.} 2007, \aap, 461, 999

\bibitem[{{Jansen}(1995)}]{jansen:thesis}
{Jansen}, D.~J. 1995, Ph.D.~Thesis, Leiden University

\bibitem[{{Jansen} {et~al.}(1994){Jansen}, {van Dishoeck}, \&
  {Black}}]{jansen:ic63}
{Jansen}, D.~J., {van Dishoeck}, E.~F., \& {Black}, J.~H. 1994, \aap, 282, 605

\bibitem[{{Johnstone} {et~al.}(2003){Johnstone}, {Boonman}, \& {van
  Dishoeck}}]{johnstone:astrochem}
{Johnstone}, D., {Boonman}, A.~M.~S., \& {van Dishoeck}, E.~F. 2003, \aap, 412,
  157

\bibitem[{{Leung} \& {Liszt}(1976)}]{leung:liszt}
{Leung}, C.-M. \& {Liszt}, H.~S. 1976, \apj, 208, 732

\bibitem[{{Leurini} {et~al.}(2004){Leurini}, {Schilke}, {Menten}, {Flower},
  {Pottage}, \& {Xu}}]{leurini:ch3oh}
{Leurini}, S., {Schilke}, P., {Menten}, K.~M., {et~al.} 2004, \aap, 422, 573

\bibitem[{{Mangum} \& {Wootten}(1993)}]{mangum:h2co}
{Mangum}, J.~G. \& {Wootten}, A. 1993, \apjs, 89, 123

\bibitem[{{Maret} {et~al.}(2005){Maret}, {Ceccarelli}, {Tielens}, {Caux},
  {Lefloch}, {Faure}, {Castets}, \& {Flower}}]{maret:meth}
{Maret}, S., {Ceccarelli}, C., {Tielens}, A.~G.~G.~M., {et~al.} 2005, \aap,
  442, 527

\bibitem[{{Mathis} {et~al.}(1983){Mathis}, {Mezger}, \&
  {Panagia}}]{mathis:isrf}
{Mathis}, J.~S., {Mezger}, P.~G., \& {Panagia}, N. 1983, \aap, 128, 212

\bibitem[{{Mihalas}(1978)}]{mihalas:book}
{Mihalas}, D. 1978, {Stellar atmospheres (2nd edition)} (San Francisco,
  W.~H.~Freeman and Co.)

\bibitem[{{M{\"u}ller} {et~al.}(2001){M{\"u}ller}, {Thorwirth}, {Roth}, \&
  {Winnewisser}}]{mueller:cdms}
{M{\"u}ller}, H.~S.~P., {Thorwirth}, S., {Roth}, D.~A., \& {Winnewisser}, G.
  2001, \aap, 370, L49

\bibitem[{{Ossenkopf} {et~al.}(2001){Ossenkopf}, {Trojan}, \&
  {Stutzki}}]{ossenkopf:radtrans}
{Ossenkopf}, V., {Trojan}, C., \& {Stutzki}, J. 2001, \aap, 378, 608

\bibitem[{{Osterbrock} \& {Ferland}(2006)}]{osterbrock:book}
{Osterbrock}, D.~E. \& {Ferland}, G.~J. 2006, {Astrophysics of gaseous nebulae
  and active galactic nuclei (2nd edition)} (University Science Books)

\bibitem[{{Poelman} \& {Spaans}(2005)}]{poelman:escprob}
{Poelman}, D.~R. \& {Spaans}, M. 2005, \aap, 440, 559

\bibitem[{{Prasad} \& {Tarafdar}(1983)}]{prasad:tarafdar}
{Prasad}, S.~S. \& {Tarafdar}, S.~P. 1983, \apj, 267, 603

\bibitem[{{Roueff} {et~al.}(2002){Roueff}, {Felenbok}, {Black}, \&
  {Gry}}]{roueff:c3}
{Roueff}, E., {Felenbok}, P., {Black}, J.~H., \& {Gry}, C. 2002, \aap, 384, 629

\bibitem[{{Rybicki} \& {Lightman}(1979)}]{rybicki:lightman}
{Rybicki}, G.~B. \& {Lightman}, A.~P. 1979, {Radiative processes in
  astrophysics} (New York, Wiley-Interscience)

\bibitem[{{Sch{\"o}ier}(2000)}]{schoeier:thesis}
{Sch{\"o}ier}, F.~L. 2000, Ph.D.~Thesis, Stockholm University

\bibitem[{{Sch{\"o}ier} {et~al.}(2002){Sch{\"o}ier}, {J{\o}rgensen}, {van
  Dishoeck}, \& {Blake}}]{schoeier:16293}
{Sch{\"o}ier}, F.~L., {J{\o}rgensen}, J.~K., {van Dishoeck}, E.~F., \& {Blake},
  G.~A. 2002, \aap, 390, 1001

\bibitem[{{Sch{\"o}ier} {et~al.}(2005){Sch{\"o}ier}, {van der Tak}, {van
  Dishoeck}, \& {Black}}]{schoeier:lamda}
{Sch{\"o}ier}, F.~L., {van der Tak}, F.~F.~S., {van Dishoeck}, E.~F., \&
  {Black}, J.~H. 2005, \aap, 432, 369

\bibitem[{Sobolev(1960)}]{sobolev:book}
Sobolev, V. 1960, Moving envelopes of stars (Harvard University Press)

\bibitem[{{Spaans} \& {van Langevelde}(1992)}]{spaans:oh-masers}
{Spaans}, M. \& {van Langevelde}, H.~J. 1992, \mnras, 258, 159

\bibitem[{{Stutzki} \& {Winnewisser}(1985)}]{stutzki:winnewisser}
{Stutzki}, J. \& {Winnewisser}, G. 1985, \aap, 144, 13

\bibitem[{{Tafalla} {et~al.}(2002){Tafalla}, {Myers}, {Caselli}, {Walmsley}, \&
  {Comito}}]{tafalla:starless}
{Tafalla}, M., {Myers}, P.~C., {Caselli}, P., {Walmsley}, C.~M., \& {Comito},
  C. 2002, \apj, 569, 815

\bibitem[{{Takahashi} \& {Uehara}(2001)}]{takahashi:h2}
{Takahashi}, J. \& {Uehara}, H. 2001, \apj, 561, 843

\bibitem[{{Takahashi} {et~al.}(1983){Takahashi}, {Silk}, \&
  {Hollenbach}}]{takahashi:h2o}
{Takahashi}, T., {Silk}, J., \& {Hollenbach}, D.~J. 1983, \apj, 275, 145

\bibitem[{{Van der Tak} {et~al.}(2005){Van der Tak}, {Neufeld}, {Yates},
  {Hogerheijde}, {Bergin}, {Sch{\"o}ier}, \& {Doty}}]{vdtak:h2o-benchmark}
{Van der Tak}, F., {Neufeld}, D., {Yates}, J., {et~al.} 2005, in The Dusty and
  Molecular Universe: A Prelude to Herschel and ALMA, ed. A.~{Wilson}, 431--432

\bibitem[{{Van der Tak}(2005)}]{vdtak:catania}
{Van der Tak}, F.~F.~S. 2005, in IAU Symposium 227: Massive Star Birth, ed.
  R.~{Cesaroni}, M.~{Felli}, E.~{Churchwell}, \& M.~{Walmsley} (Cambridge:
  University Press), 70--79

\bibitem[{{Van der Tak}(2006)}]{vdtak:london}
{Van der Tak}, F.~F.~S. 2006, Phil. Trans. R. Soc. Lond., 364, 3101

\bibitem[{{Van der Tak} {et~al.}(2000{\natexlab{a}}){Van der Tak}, {van
  Dishoeck}, \& {Caselli}}]{vdtak:ch3oh}
{Van der Tak}, F.~F.~S., {van Dishoeck}, E.~F., \& {Caselli}, P.
  2000{\natexlab{a}}, \aap, 361, 327

\bibitem[{{Van der Tak} {et~al.}(1999){Van der Tak}, {van Dishoeck}, {Evans},
  {Bakker}, \& {Blake}}]{vdtak:gl2591}
{Van der Tak}, F.~F.~S., {van Dishoeck}, E.~F., {Evans}, II, N.~J., {Bakker},
  E.~J., \& {Blake}, G.~A. 1999, \apj, 522, 991

\bibitem[{{Van der Tak} {et~al.}(2000{\natexlab{b}}){Van der Tak}, {van
  Dishoeck}, {Evans}, \& {Blake}}]{vdtak:massive}
{Van der Tak}, F.~F.~S., {van Dishoeck}, E.~F., {Evans}, II, N.~J., \& {Blake},
  G.~A. 2000{\natexlab{b}}, \apj, 537, 283

\bibitem[{{Van Dishoeck} {et~al.}(1995){Van Dishoeck}, {Blake}, {Jansen}, \&
  {Groesbeck}}]{vdishoeck:16293}
{Van Dishoeck}, E.~F., {Blake}, G.~A., {Jansen}, D.~J., \& {Groesbeck}, T.~D.
  1995, \apj, 447, 760

\bibitem[{{Van Dishoeck} \& {Hogerheijde}(1999)}]{vdishoeck:creteII}
{Van Dishoeck}, E.~F. \& {Hogerheijde}, M.~R. 1999, in NATO ASIC Proc. 540: The
  Origin of Stars and Planetary Systems, ed. C.~J. {Lada} \& N.~D. {Kylafis},
  97

\bibitem[{{Van Zadelhoff} {et~al.}(2002){Van Zadelhoff}, {Dullemond}, {van der
  Tak}, {Yates}, {Doty}, {Ossenkopf}, {Hogerheijde}, {Juvela}, {Wiesemeyer}, \&
  {Sch{\"o}ier}}]{zadelhoff:benchmark}
{Van Zadelhoff}, G.-J., {Dullemond}, C.~P., {van der Tak}, F.~F.~S., {et~al.}
  2002, \aap, 395, 373

\bibitem[{{Wright} {et~al.}(1991){Wright}, {Mather}, {Bennett}, {Cheng},
  {Shafer}, {Fixsen}, {Eplee}, {Isaacman}, {Read}, {Boggess}, {Gulkis},
  {Hauser}, {Janssen}, {Kelsall}, {Lubin}, {Meyer}, {Moseley}, {Murdock},
  {Silverberg}, {Smoot}, {Weiss}, \& {Wilkinson}}]{wright:isrf}
{Wright}, E.~L., {Mather}, J.~C., {Bennett}, C.~L., {et~al.} 1991, \apj, 381,
  200

\bibitem[{{Yates} {et~al.}(1997){Yates}, {Field}, \& {Gray}}]{yates:maser}
{Yates}, J.~A., {Field}, D., \& {Gray}, M.~D. 1997, \mnras, 285, 303

\end{thebibliography}
